\documentclass[10pt,sigconf,letterpaper,nonacm]{acmart}

\usepackage{graphicx}
\usepackage{booktabs}
\usepackage{amsmath}
\usepackage{algorithm}
\usepackage{algorithmic}
\usepackage{graphicx}
\usepackage{xspace,soul}
\usepackage{url}
\usepackage{xurl}
\usepackage{tabularx}
\usepackage{harveyballs}
\usepackage{fancyhdr}
\usepackage{comment}
\usepackage{multirow}
\usepackage{subfigure}
\usepackage[utf8]{inputenc}
\usepackage[flushleft]{threeparttable}
\usepackage{longtable}
\usepackage{pifont}
\usepackage{xcolor}
\usepackage{booktabs} 
\usepackage{enumitem} 
\usepackage{circledsteps}

\newcommand{\PaperNumber}{XXX}
\newcommand{\datadome}{DataDome\xspace}
\newcommand{\botd}{BotD\xspace}

\newcommand{\botdevasiondec}{44.95\%\xspace}
\newcommand{\datadomeevasiondec}{48.11\%\xspace}

\newcommand{\smartbots}{evasive bots\xspace}
\newcommand{\smartbot}{evasive bot\xspace}
\newcommand{\Smartbots}{Evasive bots\xspace}

\newcommand{\smart}{evasive\xspace}

\newcommand{\botserv}{bot service\xspace}
\newcommand{\botservs}{bot services\xspace}

\newcommand{\Botservs}{Bot services\xspace}

\newcommand{\antibotserv}{anti-bot service\xspace}
\newcommand{\antibotservs}{anti-bot services\xspace}

\newcommand{\Antibotservs}{Anti-bot services\xspace}

\newcommand{\para}[1]{\smallskip \noindent \textbf{#1}}

\newcommand{\system}{FP-Inconsistent\xspace}

\newcommand{\tikzxmark}{%
\tikz[scale=0.23] {
    \draw[line width=0.7,line cap=round, color=red] (0,0) to [bend left=6] (1,1);
    \draw[line width=0.7,line cap=round, color=red] (0.2,0.95) to [bend right=3] (0.8,0.05);
}}
\newcommand{\tikzcmark}{%
\tikz[scale=0.23] {
    \draw[line width=0.7,line cap=round, color=green!100] (0.25,0) to [bend left=10] (1,1);
    \draw[line width=0.8,line cap=round, color=green!100] (0,0.35) to [bend right=1] (0.23,0);
}}
\begin{document}

\title{\system: Measurement and Analysis of Fingerprint Inconsistencies in Evasive Bot Traffic}

\author{Hari Venugopalan}
\email{hvenugopalan@ucdavis.edu}
\affiliation{
\institution{UC Davis}
\country{ }
}

\author{Shaoor Munir}
\email{smunir@ucdavis.edu}
\affiliation{
\institution{UC Davis}
\country{ }
}

\author{Shuaib Ahmed}
\email{shuahmed@ucdavis.edu}
\affiliation{
\institution{UC Davis}
\country{ }
}

\author{Tangbaihe Wang}
\email{monwang@ucdavis.edu}
\affiliation{
\institution{UC Davis}
\country{ }
}

\author{Samuel T. King}
\email{kingst@ucdavis.edu}
\affiliation{
\institution{UC Davis}
\country{ }
}

\author{Zubair Shafiq}
\email{zubair@ucdavis.edu}
\affiliation{
\institution{UC Davis}
\country{ }
}

\begin{abstract}
Browser fingerprinting is used for bot detection. 
In response, bots have started altering their fingerprints to  
evade detection.
We conduct the first large-scale evaluation to study whether and how altering fingerprints helps bots evade detection.
To systematically investigate such \smartbots, we deploy a honey site that includes two \antibotservs (\datadome and \botd) and solicit bot traffic from 20 different \botservs that purport to sell ``realistic and undetectable traffic.'' 
Across half a million requests recorded on our honey site, we find an average evasion rate of 52.93\% against \datadome and 44.56\% evasion rate against \botd. 
Our analysis of fingerprint attributes of \smartbots shows that they indeed alter their fingerprints.
Moreover, we find that the attributes of these altered fingerprints are often inconsistent with each other.
We propose \system, a data-driven approach to detect such inconsistencies across \textit{space} (two attributes in a given browser fingerprint) and \textit{time} (a single attribute at two different points in time).
Our evaluation shows that our approach can reduce the evasion rate of \smartbots by \botdevasiondec-\datadomeevasiondec while maintaining a true negative rate of 96.84\% on traffic from real users.

\end{abstract}

\maketitle
\section{Introduction}
The prevalence of bots on the web is on the rise~\cite{imperva_evasive}.
Per recent reports, bots constitute around 49.6\% of online traffic~\cite{imperva_2024}, with 64.5\% of those being bots that engage in malicious activity.
Bad actors employ bots to launch a multitude of attacks~\cite{impression-fraud, viceroi, boxer, csci_casestudy, industrial_scraping}.
To counter such attacks, \antibotservs aim to detect and block bot traffic.
Prior research has shown that \antibotservs use browser fingerprinting to detect bots without disrupting the user experience of legitimate users~\cite{Vastel2020MADWeb, double-edged-sword}.
Browser fingerprints capture attributes of the web browser sending web requests and \antibotservs attempt to use differences in these attributes to distinguish bots from real users~\cite{him-many-faces}.

Blackhat marketplaces~\cite{babylon, seoclerks, spark}, however, advertise realistic and undetectable bot traffic as a service.
The traffic from such services constitute impression fraud and serve to artificially boost website engagement for monetization~\cite{impression-fraud,farooqi2017characterizing,javed2015measurement}.
To evade detection, bots from these services are likely altering their fingerprint attributes that are used by \antibotservs for detection~\cite{hlisa, reliable-web-crawling}.
We refer to such bots as \emph{\smart} bots.
It is important to characterize \smartbots and their fingerprints to improve the effectiveness of bot detection.

Prior research has studied bot fingerprints by employing their own bots~\cite{webrunner-2049, Vastel2020MADWeb} or studying naturally discovered bots on their honey sites~\cite{Li2021S&P}.
Thus, this line of work is not geared towards capturing the \smart fingerprints used by bots seeking to evade detection in the wild.
Wu et al. performed a large-scale characterization of the differences between human and bot fingerprints in the wild~\cite{him-many-faces}.
However, they did not specifically characterize \smartbots since they treat their bot detection system decisions as ground-truth to distinguish between the fingerprints of bots and real users. 

To fill this gap, we perform the first large-scale measurement of \smartbots that evade \antibotservs.
To this end, we drive traffic from different \botservs from blackhat marketplaces to different instances of our honey site.
That way, the requests recorded at each honey site instance can be attributed to a \botserv from whom we purchased traffic.
These operators advertise their traffic as being realistic and natural, indicating that they likely employ \smartbots to ensure that they do not get detected.
We integrate two commercial bot detection services (\datadome and \botd) on our honey site for bot detection.
We also instrument our honey site to collect a range of fingerprint attributes.

%

We collect 507,080 requests from 20 different \botservs, with \datadome and \botd detecting 55.44\% and 47.07\% of these requests respectively.
We analyze fingerprint attributes from different \botservs to identify different sets of attributes that are effective at evading \datadome and \botd individually as well as attributes that are effective at evading both \antibotservs.
Our analysis reveals spatial inconsistencies (among the different attributes of a given request) and temporal inconsistencies (across requests originating from the same device).
These include obvious inconsistencies that cannot exist for real users, thereby making them useful signatures to detect bots.


We use observations from our analysis to develop \system, a data-driven approach to discover inconsistencies in fingerprint attributes for bot detection.
\system relies on the insight that real devices can only have a limited number of hardware and software configurations that are reflected in fingerprint attributes.
\Smartbots, in their attempt to evade detection, emulate a large number of invalid or extraneous configurations.
\system leverages this mismatch between the expected and observed number of configurations to identify potential inconsistencies among \smartbot fingerprints.
It does so by calculating the number of configurations for pairs of fingerprint attributes from \smartbots and identifying inconsistencies among attribute pairs that exhibit a higher-than-expected number of configurations.

Using this approach, we generate inconsistency rules that can be readily deployed by \antibotservs.
Prior research focusing on the use of inconsistencies for bot detection \cite{Vastel2018SP, Vastel2020MADWeb} has
predominantly relied on one-off anecdotes to define inconsistencies that are not data-driven and hence do not scale. 
%
\system systematizes the generation of inconsistency rules for bot detection.

Our evaluation shows the rules generated by \system are able to achieve \datadomeevasiondec and \botdevasiondec reduction in traffic that evades \datadome and \botd respectively while maintaining a true negative rate of 96.84\% on real user traffic. Our experiments also show that \system does not incur false positives with
most privacy-enhancing technologies.
We open-source our honey site architecture and inconsistency rules for public use \href{https://github.com/hariv/fp_inconsistent}{at this link}.

Our work makes the following contributions:
\begin{itemize}

    \item A \textbf{novel honey site architecture to establish reliable ground-truth} for \smart bots.
    
    \item A \textbf{large-scale measurement and analysis of fingerprint attributes} for \smart bots that are able to evade \antibotservs.
    
    \item A \textbf{data-driven approach to discover inconsistencies in fingerprint attributes} for detecting \smartbots.

\end{itemize}
\section{Background and Related Work}
\subsection{Evaluation of bot detection services}
\Antibotservs on the web generally employ machine learning to determine if an incoming request was sent by a human or a bot\cite{Cabri2018HPCC}.
These services rely on several signals captured through different browser fingerprinting APIs, request headers, 
and behavior characteristics on a website\cite{Vastel2020MADWeb}.
Prior research has attempted to measure the accuracy of \antibotservs and understand their detection techniques.
Azad et al. \cite{webrunner-2049} analyzed 15 different \antibotservs, 14 of which used modern fingerprinting 
techniques such as WebGL and Canvas-based fingerprinting. While these services rely on inconsistencies in
fingerprint attributes to detect bots, we show how they can be more extensive in using them to improve bot
detection (Section \ref{sec:technique}). 

Azad et al.  also tried to evaluate the performance 
of these services by deploying their own bots and measuring their evasiveness. In contrast, we evaluate 
\antibotservs on requests from bots in the wild.

\subsection{Analysis of bot traffic in the wild}
Xigao et al. \cite{Li2021S&P} studied the prevalence of "malicious" bots in the wild.
They use the behavior of bots (indulging in credential stuffing, not honoring bots.txt, etc) to characterize 
them as malicious. Such characterization is not applicable for bots indulging in impression fraud since these 
bots don't exhibit any explicit behavior that can be leveraged for detection. Further, their approach draws
traffic from bots in general and they do not include mechanisms to isolate \smartbots visiting their honey sites
that seek to evade detection.

Wu et al. \cite{him-many-faces} analyzed browser fingerprints from 36 billion requests on 14 commercial websites.
Their analysis shows that adversarial bots (bots that change their fingerprints to avoid detection) have significantly different properties compared to benign bots.
While they conducted the largest study (at the time of writing) of bots in the wild, their ground-truth relies on decisions by F5 Inc.\cite{f5_bot_defense}, 
a commercial \antibotserv.
Thus, without a more robust mechanism to collect ground-truth, their approach cannot analyze bots that can evade commercial \antibotservs such as F5.

Browser Polygraph \cite{Kalantari2024BrowserPolygraph} employs machine learning to detect bots that indulge in account takeover 
fraud (ATO). They predict if the fingerprint attributes in a request are consistent with the request's reported \texttt{User-Agent}. 
In contrast, our work proposes a data-driven and semi-automatic technique to discover inconsistencies between any pair of fingerprint attributes
(which includes but is not limited to the \texttt{User-Agent}) to combat impression fraud. Further, similar to the work of Wu et. al,
their approach could be bolstered with more robust ground-truth since they rely on tags from FinOrg (a financial organization) to
provide ground-truth for evaluation. In our work, our novel honey site architecture provides ground-truth to isolate traffic
sent from different \botservs.

\subsection{Challenges in bot detection}
We discuss some of the common techniques used by bots to evade detection.

\para{Polymorphism:} Certain bots morph their User-Agent or other attributes (i.e., fingerprints) to appear as benign website visitors 
for evasion \cite{Kalantari2024BrowserPolygraph, webrunner-2049}.
Iliou et. al \cite{Iliou2019ARES} showed that while machine learning algorithms can detect simple bots with a precision and recall 
of 95\% and 97\% respectively, more advanced bots, i.e. bots that change their fingerprints, result in a drop in accuracy to only 55\%.

\para{Behavioral Mimicry:} Bots also simulate human-like behavior to evade behavioral analysis systems, including mimicking mouse 
movements, keystrokes, browsing patterns, and human text input\cite{twitterbot}.
Bot detection systems use these movements as ``Human Interactive Proofs (HIPs)''\cite{Gianvecchio2008USENIX, Gianvecchio2009CCS} to determine 
if a website visitor is a bot or a human.
Jing et. al. \cite{Jin2013IEEE} developed a bot framework for bots to generate keystrokes and mouse clicks that closely resemble human
actions to evade detection.
\section{Threat Model}
In this paper, we focus on bots committing
impression fraud~\cite{impression-fraud}. Web publishers
who seek to artificially inflate the engagement on their 
websites indulge in this type of fraud. Inflating engagement
allows these publishers to monetize and profit from their
websites through ads, even when they cannot guarantee
visits to their website from legitimate users. Advertisers
pay publishers for impressions of their ad (views, clicks, etc) 
on the publisher's website. However, only impressions recorded
from legitimate users are useful to advertisers. Publishers who 
do not receive traffic from legitimate users could employ bots 
to record these impressions to get paid by advertisers without 
delivering any useful impressions to them. We focus on bots 
indulging in impression fraud over other types of fraud 
(such as credential stuffing, account takeover, etc), since 
these bots do not have a need to perform specific actions \cite{webrunner-2049, Li2021S&P}
to reach their goal, thereby making it more challenging to
detect them.

In our threat model, we consider publishers who incorporate \antibotservs 
on their websites to provide assurance of traffic from legitimate 
users, while employing \smartbots to evade detection.
\section{Measurement infrastructure}
In this section, we describe our measurement infrastructure including 
the design of our novel honey site architecture. We design our measurement
infrastructure to satisfy three requirements that enable us to reliably 
characterize \smartbots: first, we need reliable ground-truth that
we only record requests from \smartbots of interest and no other
entities (real users or other bots). Second, we need decisions from bot detection 
services on each request to isolate requests that evade detection.
Third, we need to collect browser attributes that constitute browser 
fingerprints in these requests to analyze attributes that help with evasion.

\subsection{Honey site architecture}
\begin{figure}
    \centering
    \includegraphics[width=\columnwidth]{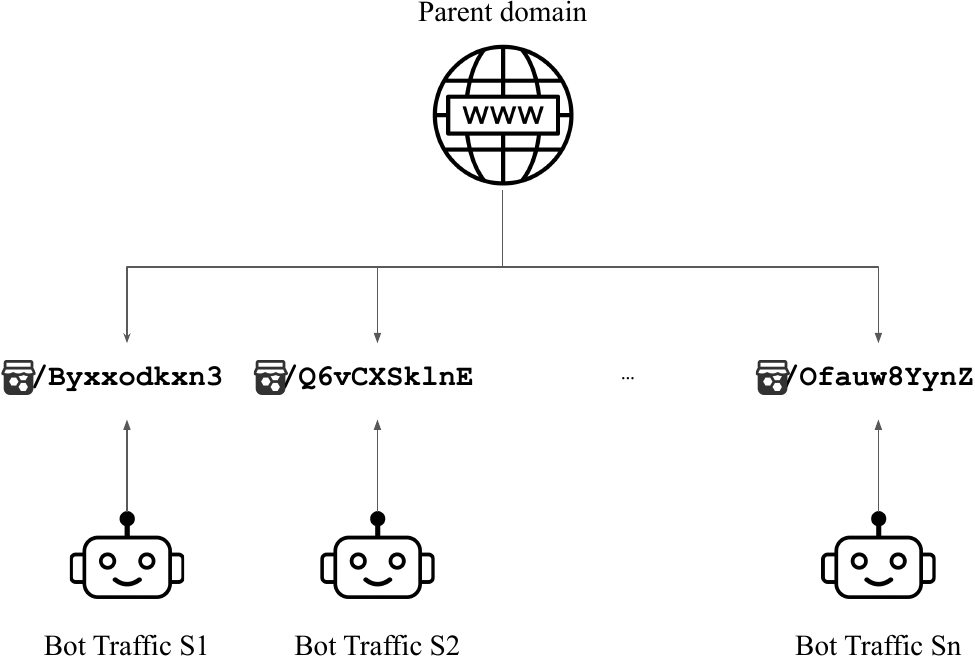}
    \caption{To collect requests from different \botservs, we create multiple versions
    of the same honey site under the same domain. The only difference between these versions
    is the presence of different random strings in their URL. We then drive traffic from different \botservs to different versions of the honey site.}
    \label{fig:honey_site_configuration}
\end{figure}

Using obscure domain names for honey sites ~\cite{Li2021S&P} cannot guarantee that the 
honey sites only receive requests from \smartbots. Bots that automatically send requests 
to such domains are typically indexing bots that visit new websites added to domain 
registries and other sources of DNS records. Examples of such bots include search engine 
bots that do not have a need to conceal their identities. In fact, Google's bots announce 
their identity through their User-Agent~\cite{google-bots}. While \smartbots may also send 
requests to such domains, the absence of a mechanism to isolate those requests makes it 
challenging to analyze them. \Smartbots indulging in impression fraud do not have a need to 
perform specific actions to record views or impressions. Hence, such bots cannot be detected 
based on their actions/behavior \cite{webrunner-2049, Li2021S&P}.

To overcome the challenge of only recording requests from \smartbots, we deploy multiple
versions of the same honey site under the same domain. These versions only differ in terms 
of the presence of arbitrarily chosen strings in their URL. We do not record requests that 
do not contain one of these strings in the URL to ensure that we do not record requests 
from real users or generic bots that discover our domain. We also share URLs having
different arbitrary strings with different \botservs. Thus, these URL strings enable
the isolation of requests received from different \botservs. As a concrete example, 
if \textit{example.com} is the domain of our honey site, \textit{example.com/XXXXX}, 
\textit{example.com/YYYYY}, and \textit{example.com/ZZZZZ} would constitute different 
versions of the honey site. We then purchase traffic from 3 different \botservs
to each send requests to one of these URLs. Real users and other generic bots who may 
stumble upon our site, will not know these strings and hence cannot include such strings
in their requests. Thus, we can ensure that we only record requests from the \botservs 
where we made our purchases using these URL strings. Figure \ref{fig:honey_site_configuration} 
shows an overview of the honey site architecture.

\subsection{\Antibotservs}
\label{subsec:antibot-services}
We integrated two popular commercial \antibotservs on our honey site: \datadome \cite{datadome} and \botd \cite{botd}.
Both \datadome and \botd provide real-time decisions on requests received 
on a website.
\datadome advertises real-time decisions for a request in under 2 
milliseconds with an overall accuracy of 99\% and a false positive rate 
of 0.01\%. Prior research on bot detection has explored \datadome \cite{webrunner-2049, datadome_citation_paper}. 
\botd is a bot detection service from the developers of the popular open-source 
fingerprinting library FingerprintJS \cite{fingerprintjs} that is widely used
in industry and academia \cite{Kalantari2024BrowserPolygraph, Li2021S&P, webrunner-2049, Vastel2020MADWeb}.
\botd claims to use ``the most advanced device fingerprinting technology'', 
and reports a detection accuracy of 99.5\%. 

We integrate JavaScript libraries of both these services on our honey site\footnote{
As required by \datadome, for each request, we also make an API call from
our server to get their decision}.
These libraries collect browser fingerprints of the browser visiting the honey site and relay them to their own servers.
The servers then respond with the decision of whether a real human or a bot originated the request.

These services are black-boxed and do not provide information on fingerprint attributes
they use as features to decide if a request originates from a bot. To determine this
information, we crawl our honey site using OpenWPM \cite{openwpm}. OpenWPM is an open-source 
tool to track the behavior of different web elements, including scripts, on a webpage.

Table \ref{tab:api_comparison_botd_datadome} in Appendix \ref{app:api-comparison} 
highlights the different browser APIs accessed by \datadome and \botd.
Both services access a number of fingerprinting APIs such as \texttt{navigator.plugins}, 
\texttt{HTMLCanvasElement.getContext}, \texttt{navigator.userAgent}, 
and more.
We find that \datadome collects more attributes from each request than \botd.
In Section \ref{sec:measurements}, we see that \datadome has higher bot detection accuracy than
\botd, which could potentially stem from these additional attributes.

\begin{figure}
    \centering
    \includegraphics[width=\linewidth]{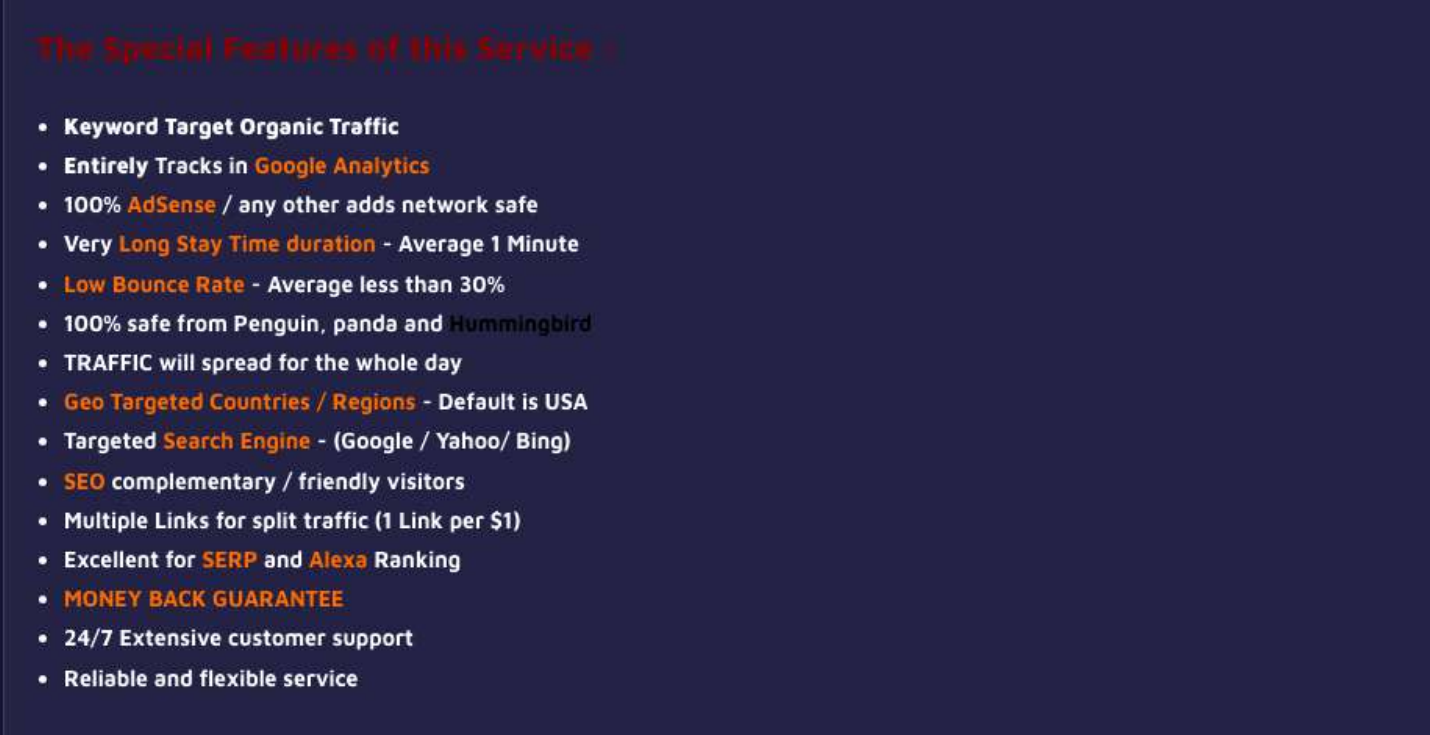}
    \caption{Screenshot from a \botserv on SEOClerks making claims about
    sending organic traffic to drive engagement on websites. The claims
    likely suggest that the \botserv employs \smartbots to took real users.}
    \label{fig:bot_claims}
\end{figure}
\subsection{\Botservs}
We made purchases from multiple online \botservs to send traffic to different versions 
of our honey site. We make our purchases from the SEOClerks~\cite{seoclerks}, an underground 
marketplace for web traffic where \botservs advertise their traffic as being real, organic, and 
Adsense safe to boost website engagement. Their claims of being able to send real and organic 
traffic indicate that they are likely using \smartbots that alter their fingerprints to look
like real users. Figure~\ref{fig:bot_claims} captures a screenshot from a \botserv on 
SEOClerks making such claims about their traffic. We share URLs with different version strings 
with different \botservs to identify the \botservs of each request on our honey site.

\subsection{Data Collection}
\label{sec:data_collected}
\begin{figure}
    \centering
    \includegraphics[width=\linewidth]{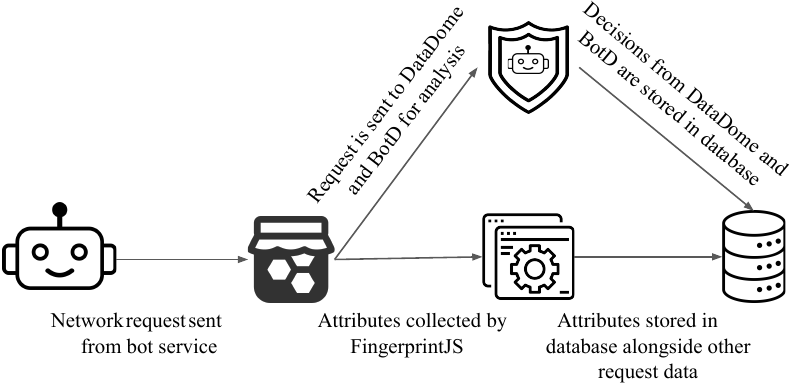}
    \caption{Overview of our data collection pipeline.}
    \label{fig:datacollection_pipeline}
\end{figure}

To characterize the differences in the fingerprint attributes
of \smartbots, we extract information from different browser APIs 
and properties upon loading our honey site in the browser. We 
send this information to our server in an http request. We use 
FingerprintJS~\cite{fingerprintjs}, a widely deployed browser 
fingerprinting library to capture this information. FingerprintJS 
captures over 30 different fingerprint attributes including the list of 
fonts installed on the browser, the number of CPU cores on the device 
running the browser, the amount of memory on the device, and 
the languages supported by the browser. 
While we focus on the attributes captured by
FingerprintJS in this paper, both our measurement
analysis (Section \ref{sec:measurements}) and our methodology to discover 
inconsistencies (Section \ref{sec:technique}) 
are compatible with other fingerprint attributes
too.
\section{Analysis}
\label{sec:measurements}
\begin{table}[!htpb]
    \centering
    \caption{Overview of different bot services sending traffic to our honey site and their evasion rates against \datadome and \botd.}
        \small
          \resizebox*{0.8\linewidth}{!}{%
            \begin{tabular}{l c c c}
            \toprule
        {\bf{Bot}} & {\bf{Num.}} & {\bf{DataDome}} & {\bf{BotD}}\\
        {\bf{Service}} & {\bf{Requests}} &  {\bf{Evasion Rate}} & {\bf{Evasion Rate}}\\
        \midrule
        {S1} & {121500} & {44.01\%} & {71.58\%} \\
        {S2} & {63708} & {42.99\%} & {72.29\%} \\
        {S3} & {54746} & {74.91\%} & {10.26\%} \\ 
        {S4} & {47278} & {38.65\%} & {73.85\%} \\
        {S5} & {40087} & {23.86\%} & {72.65\%} \\
        {S6} & {32447} & {71.81\%} & {5.45\%} \\
        {S7} & {28940} & {2.56\%} & {39.99\%} \\
        {S8} & {26335} & {80.43\%} & {28.9\%} \\
        {S9} & {23412} & {78.29\%} & {19.33\%} \\
        {S10} & {18967} & {15.77\%} & {59.23\%} \\
        {S11} & {17996} & {6.55\%} & {59.36\%} \\
        {S12} & {7010} & {5.05\%} & {51.44\%} \\
        {S13} & {5119} & {6.95\%} & {50.52\%} \\
        {S14} & {4920} & {83.74\%} & {90.08\%} \\
        {S15} & {4219} & {11.14\%} & {100\%} \\
        {S16} & {4174} & {4.48\%} & {0.02\%} \\
        {S17} & {2999} & {74.66\%} & {7.9\%} \\
        {S18} & {1430} & {20.7\%} & {100\%} \\
        {S19} & {1411} & {9.92\%} & {100\%} \\
        {S20} & {382} & {97.12\%} & {97.12\%} \\
    \bottomrule
     \end{tabular}
     }
     \label{tab:service_stats}
     \end{table}
Over a period of 3 months, from September 2023 to November 2023, we received
507,080 requests from 20 different \botservs.
We first report the detection rate of the \antibotservs 
and then compare fingerprint attributes of bots that evade detection
against those that were detected. This analysis helps understand the
attributes used by bots for evasion and ways to overcome them.

Table \ref{tab:service_stats} shows the statistics of the traffic obtained from each 
\botserv along with the evasion rate against
the two \antibotservs on our honey site (\datadome and \botd).
Among the 507,080 requests we received, 55.44\% of requests were detected by 
\datadome, and 47.07\% of requests were detected by \botd. These results show 
that a significant proportion of bots are able to evade \antibotservs. 
\newline\newline
\fbox{\begin{minipage}{0.96\columnwidth}\textbf{Takeaway 1:} Our measurement shows
that \smartbots are not reliably detected by commercial \antibotservs.
\end{minipage}
\label{takeaway:elusive_bots}
}
\subsection{IP addresses for evasion}
We observed requests on our honey site that contained IP addresses with 
Autonomous System Numbers (ASNs) mapping to cloud services such as Amazon 
Web Services (AWS). Since such ASNs are likely flagged as those used by 
bots~\cite{fastflux, rbseeker}, we check the ASNs of the requests we 
received against public ASN block lists~\cite{brianhama, growtoups}.
We report that 82.54\% of requests originated from flagged ASNs. 
Among these, 52.93\% of requests evade \botd and 43.17\% of requests 
evade \botd. These results show that \smartbots are able to 
evade detection even when they send requests from flagged ASNs.

We suspect that \antibotservs may not rely on ASN block lists since 
real users and bots can share the same ASNs but can send requests 
from different IP addresses. Accordingly, we ran similar analysis 
with blocked IP addresses using MaxMind's minFraud API \cite{maxmind-minfraud}. 
Consistent with findings in prior research~\cite{Li2021S&P}, we find that 
IP block lists offer limited coverage (15.86\%). More interestingly, among the IP 
addresses that were covered, requests from 48.1\% were able to evade \datadome 
and 68.85\% were able to evade \botd.

In conclusion, we see that a significant number of bots that sent requests 
from blocked IP addresses and ASNs were able to evade both \datadome and
\botd. This indicates that \smartbots don't merely send requests from IP 
addresses not captured by block lists to evade detection.
\newline\newline
\fbox{\begin{minipage}{0.96\columnwidth}\textbf{Takeaway 2:} \Smartbots do not merely rely 
on sending requests from IP addresses that are not captured by block lists to 
evade detection.
\end{minipage}}

\subsection{Fingerprint attributes for evasion}
\label{sec:measurement_attributes}
Since \smartbots don't merely rely on IP addresses, we systematically analyze the browser
fingerprint attributes in their requests to identify those used for evasion. Concretely, we 
train models to distinguish between the requests that were detected by and evaded \datadome 
and \botd respectively. We then use techniques from the explainability of machine learning to identify 
fingerprint attribute values that help with evasion. We then explore the values of these 
attributes on requests from \botservs that were most successful with evasion to verify that 
they enable evasion.

\subsubsection{Identifying fingerprint attributes} We train two random forest classifiers 
using XGBoost~\cite{xgboost} to distinguish between the requests that were detected and 
evaded \datadome and \botd respectively. Each classifier takes as input fingerprint attributes 
from each request (discussed in Section \ref{sec:data_collected}) and provides a binary decision 
on whether that request would detected by the respective \antibotserv.

We performed a 90-10 split on the requests to train the classifiers. The classifier for 
\botd attained an accuracy 97.8\% on the training set and 97.71\% on the test set while the classifier 
for \datadome attained an accuracy of 82.09\% on the training set and 81.66\% on the test set. 
These high accuracy values indicate that the fingerprint attributes of requests that evade the 
two \antibotservs are considerably different from those of requests detected by them.

\begin{table}[h]
\centering
  \caption{Top 5 most important fingerprint attributes that help evade \datadome and \botd.}
  \label{tab:shap_features}
\begin{tabular}{c c}
  \toprule
  \textbf{\datadome} & \textbf{\botd} \\
  \midrule
  {Vendor Flavors} & {Vendor Flavors} \\
  {Plugins} & {Plugins} \\
  {Screen Frame} & {Touch Support} \\
  {Hardware Concurrency} & {Vendor} \\
  {Forced Colors} & {Contrast} \\
  \bottomrule
\end{tabular}
\end{table}

We use SHapley Additive exPlanations or SHAP~\cite{shap} to analyze these classifiers to 
identify fingerprint attributes that result in evasion. Table Table~\ref{tab:shap_features} 
lists the top 5 attributes that help evade \datadome and \botd respectively.

\subsection{Fingerprint attributes among \smartbots}
We now inspect the attribute values of requests from \botservs with high 
evasion rates to see if they exploit the attributes identified in Table~\ref{tab:shap_features}
 for evasion. Concretely, we compare attribute values across \botservs 
that have high evasion rates against those that have low evasion rates.

\begin{figure}[h]
\centering
  \includegraphics[width=0.9\columnwidth]{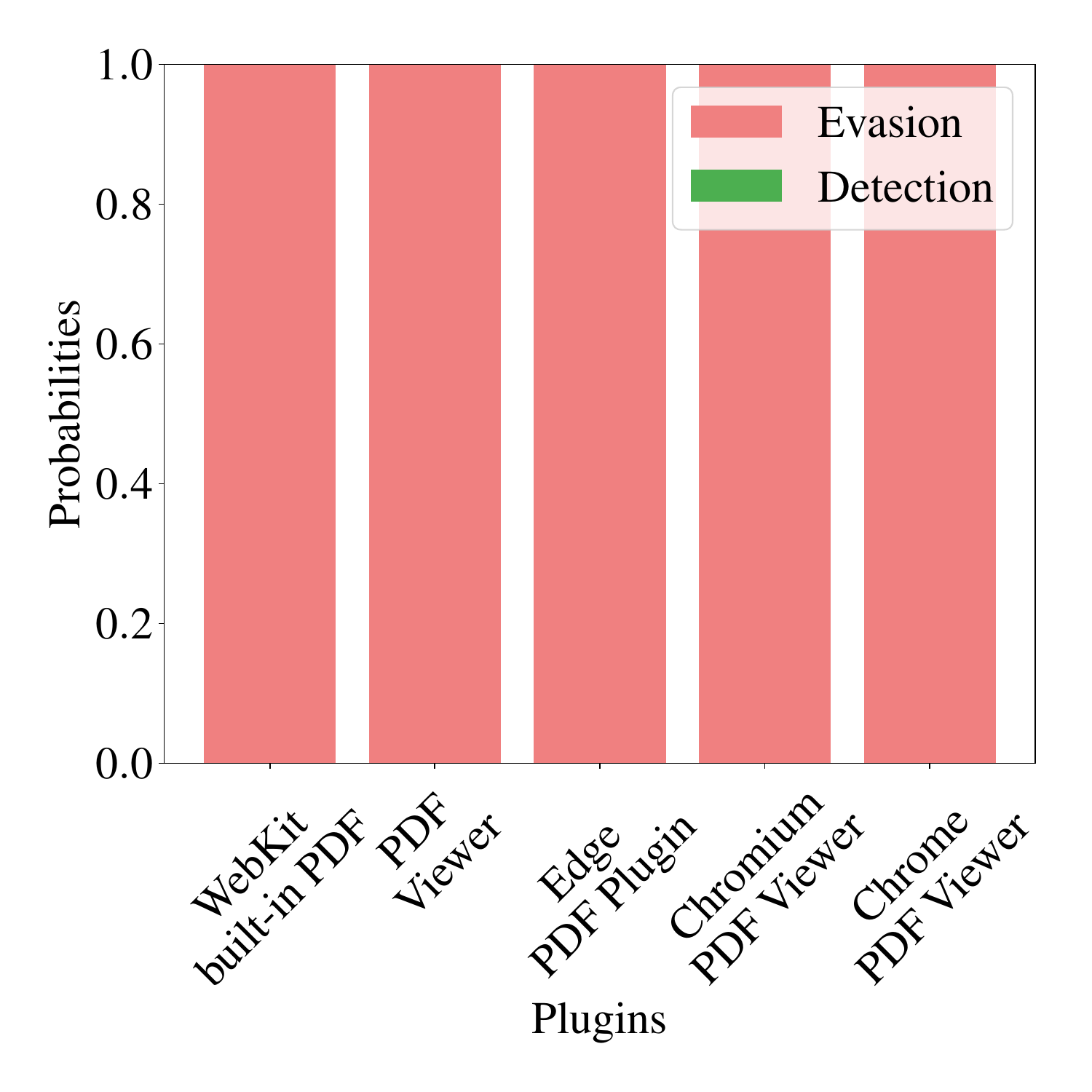}
  \caption{Bar plot showing the probability of PDF plugins that have the highest 
  probability of evasion against \botd. This plot shows that the presence of any 
  plugin helps evade \botd.}
  \label{fig:botd_plugins}
\end{figure}

\subsubsection{Bots evading \botd}
\label{sec:botd_evade}
We inspected requests from the top 3
\botservs with the highest evasion rates against \botd (S15, S18, and S19 in Table \ref{tab:service_stats}) and
the top 3 \botservs with the lowest evasion rates against \botd (S6, S16, and S17 in \ref{tab:service_stats}). We record
7,132 requests from the top 3 \botservs evading \botd and report 100\% evasion
among them. We record 39,620 requests from the top 3 \botservs that are 
detected by \botd and report an evasion rate of 5.11\% among them. 

We did not observe significant differences between the values of \texttt{Vendor Flavors}, 
\texttt{Vendor}, and \texttt{Touch Support} attributes among requests from these \botservs.
99.91\% of requests from services evading \botd supported the Chrome PDF Viewer plugin, 
while 100\% of requests detected by \botd did not support any plugins. Motivated by these 
stark differences, we further investigate the impact of plugins on evading \botd. Concretely, 
from all requests received on our honey site, we compute the probability of evading \botd 
when supporting any one of 5 commonly used PDF plugins. Figure~\ref{fig:botd_plugins} shows that 
the presence of any PDF plugin nearly guarantees evasion against \botd.

\subsubsection{Bots evading \datadome} 
\label{sec:dd_evade}
We similarly inspect requests from
the top 3 \botservs with the highest and lowest evasion rates against \datadome. We 
record 52,746 requests from the top 3 \botservs
evading \datadome (S8, S9, and S17 in Table \ref{tab:service_stats}) having 79.15\% evasion among them. We 
record 51,110 requests from the top 3 \botservs 
detected by \datadome (S7, S11, and S16 in Table \ref{tab:service_stats}) with an evasion rate of 4.12\%. 

\begin{figure}[h]
\centering
  \includegraphics[width=0.9\columnwidth]{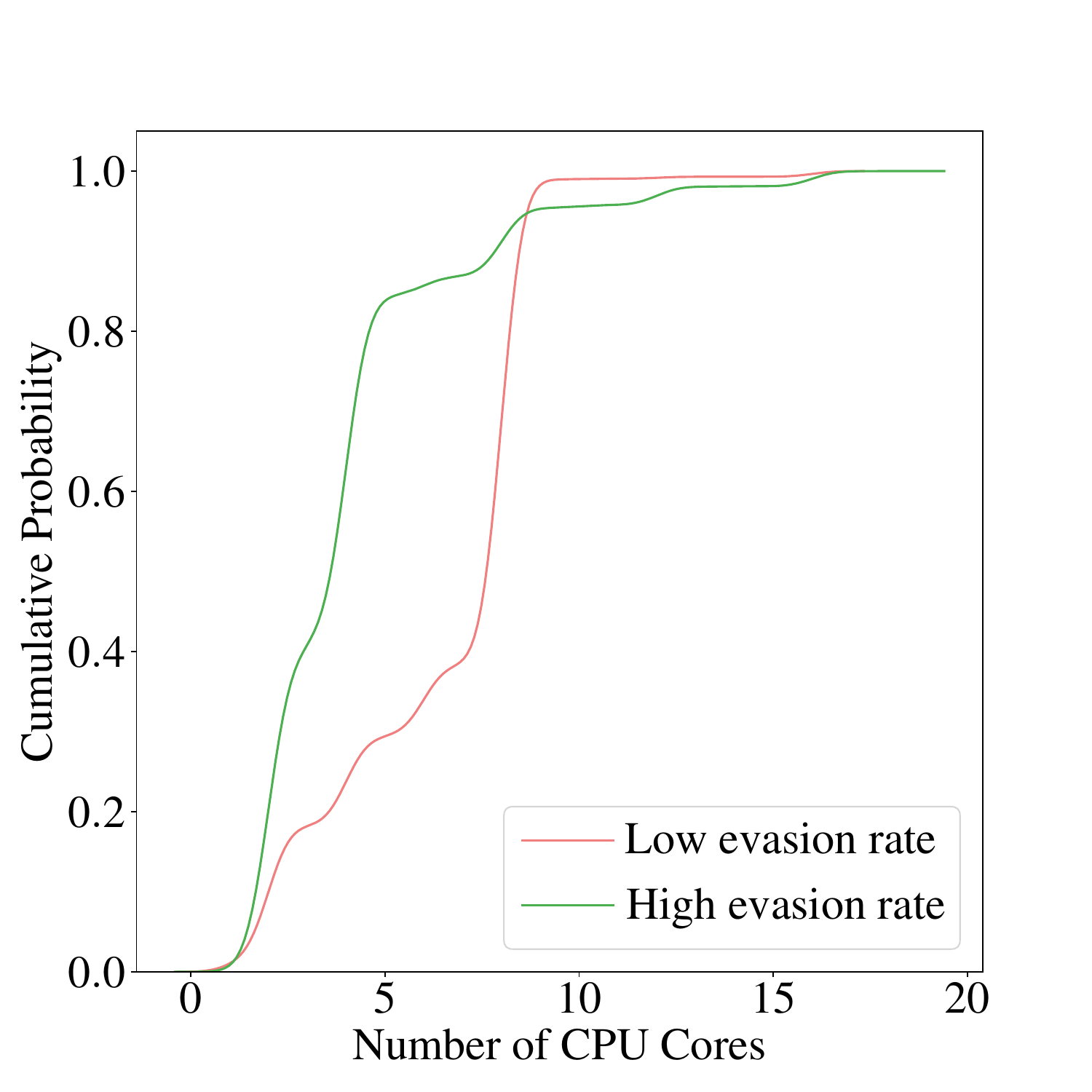}
  \caption{Cumulative probability distribution function (CDF) plots
          of the number of CPU cores recorded on requests from \botservs
          that had the highest evasion rate over \datadome against those
          that had the the lowest evasion rate over \datadome.}
  \label{fig:cdf_conc_dd}
\end{figure}

100\% of requests from the top 3 \botservs having the 
highest evasion rate against \datadome did not support 
any plugins.
However, 56.45\% of requests from the 3 \botservs with the 
lowest evasion rate against \datadome
did not support any plugins either. Analyzing the \texttt{Screen Frame} 
and \texttt{Forced Colors} attributes revealed certain values 
that always result in detection. However, we did not observe 
values for these attributes that help with evasion.  

Figure~\ref{fig:cdf_conc_dd} compares cumulative probability
distribution functions (CDFs) of the number of CPU
cores (captured by \texttt{hardwareConcurrency}) on requests from 
\botservs with high evasion rates over \datadome against 
the values on requests from \botservs with low evasion rates 
over \datadome. These results indicate that low values for
\texttt{hardwareConcurrency} help evade \datadome. Concretely, 
84.7\% of requests from \botservs with a high evasion rate against 
\datadome had fewer than 8 cores. In contrast, only 38.16\% of requests 
from \botservs detected by \datadome had fewer than 8 cores. To further
assess the impact of \texttt{hardwareConcurrency}, we disregard requests 
that contain values for \texttt{Screen Frame} and \texttt{Forced Colors}
that always lead to evasion. Now, 84.7\% of requests from \botservs 
with a high evasion rate against \datadome have fewer than 8 
cores while only 19.05\% of requests from \botservs with a low evasion 
rate against \datadome have fewer than 8 cores. 

\Smartbots evading \datadome ensure certain values for combinations of attributes. 
This is different from \smartbots evading \botd that ensured certain values for 
one set of attributes (plugins). We investigate more combinations of attributes 
that help evade \datadome in Appendix \ref{app:datadome_tree}.

\subsubsection{Bots evading \datadome and \botd}
Requests from two different \botservs have over 80\% evasion rate against
both \datadome and \botd (S14 and S20 in Table \ref{tab:service_stats}). We received 
5,302 requests from these services which have an 84.7\% evasion rate against
\datadome and 90.59\% evasion against \botd.

We observe that 83.77\% of these requests have fewer than 8 CPU
cores indicating that they exploit hardware concurrency to evade
\datadome. Interestingly, 93.02\% of these requests do not have
any plugins, indicating that they do not exploit plugins to evade
\botd. They exploit \texttt{touchSupport}, a different blind spot 
of \botd for evasion. Concretely, 78.36\% of requests from the 
\botservs evading both \datadome and \botd support touch events, 
while only 3.95\% of requests from the top 3 \botservs
having the lowest evasion rate against \botd support touch events.
In contrast, only 0.07\% of requests from the top 3 \botservs that 
only evaded \botd (Section \ref{sec:botd_evade}) showed support 
for touch events and 8.61\% of requests from the top 3 \botservs 
that only evaded \datadome (Section \ref{sec:dd_evade}) showed 
support for touch events.
\newline\newline
\fbox{\begin{minipage}{0.96\columnwidth}\textbf{Takeaway 3:} \Smartbots exploit either
\texttt{touchSupport} and \texttt{plugins} to evade \botd. They exploit 
\texttt{hardwareConcurrency} to evade \datadome.
\end{minipage}
\label{takeaway:combined}
}
\section{Inconsistency analysis}
\label{sec:inconsistency_measurement}
From our analysis in the previous section, we see ensuring certain values for certain
fingerprint attributes helps bots evade detection.
One way in which \smartbots could accomplish this would be to send requests from devices that would contain the desired values for attributes.
For example, \smartbots could send requests from devices containing 4 CPU cores to ensure a value of 4 for \texttt{hardwareConcurrency}.
Alternatively, \smartbots could alter browser APIs and device properties to present their desired values for fingerprint attributes \cite{liu2022gummy}.
In this case, an \smartbot could alter the \texttt{hardwareConcurrency} attribute of the \texttt{navigator} object to return 4 on a device that may not have 4 CPU cores.

In this section, we describe various inconsistencies in fingerprint attributes among the requests received on our honey site.
These inconsistencies provide evidence of bots altering browser APIs since such inconsistencies are 
extremely unlikely to occur when using real devices.
%
%
%
%
We use insights from these inconsistencies to develop \system, our semi-automated technique to generate inconsistency rules to detect 
\smartbots (Section~\ref{sec:technique}).

\begin{figure}[b]
\centering
  \includegraphics[width=0.9\columnwidth]{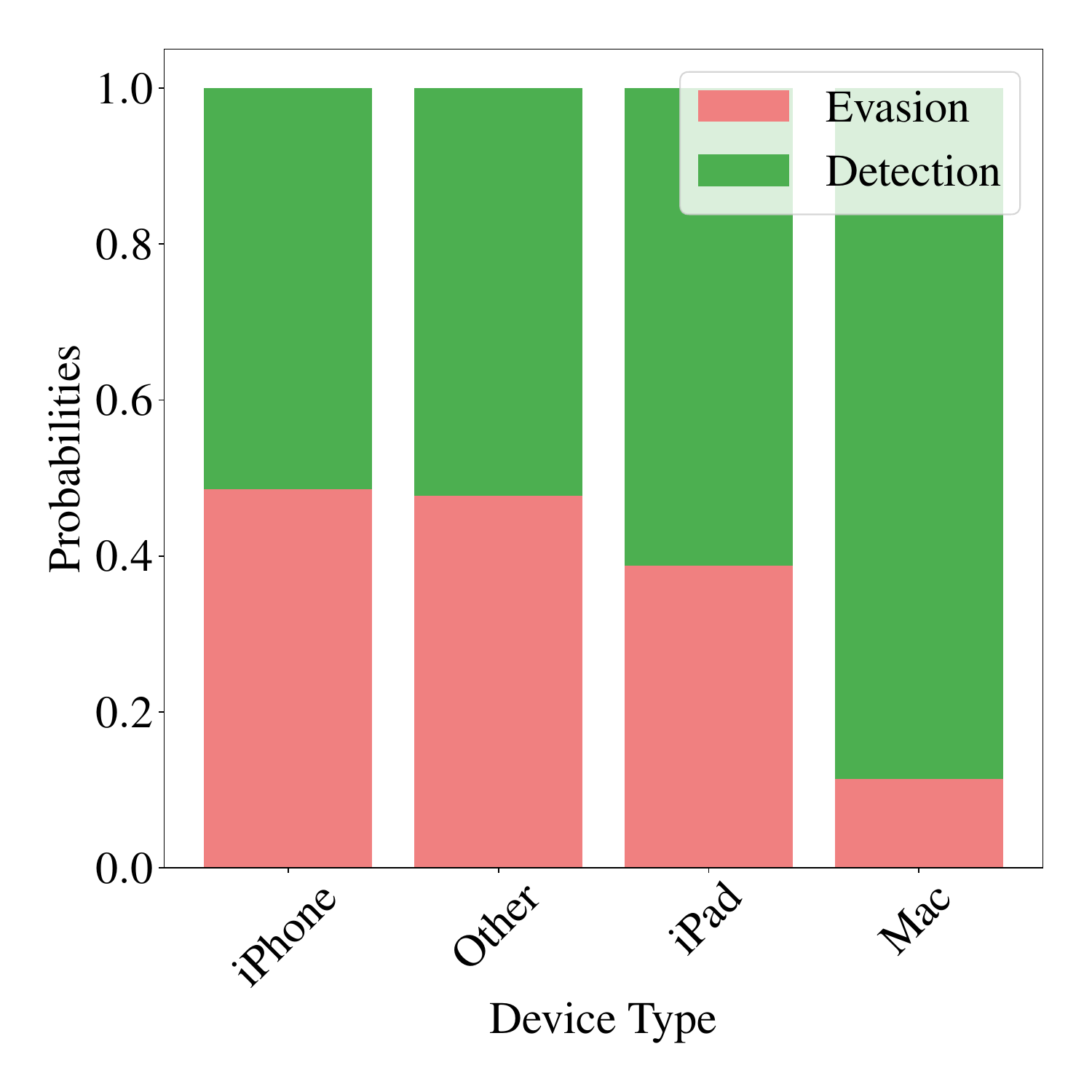}
  \caption{Bar plot showing the top 4 device types (inferred from the User-Agent)
  that have the highest probability of evading \datadome.}
  \label{fig:inverse_device_type}
\end{figure}
\subsection{Inconsistencies across fingerprint attributes}
\label{sec:other_spatial}
\begin{figure}[b]
\centering
  \includegraphics[width=0.9\columnwidth]{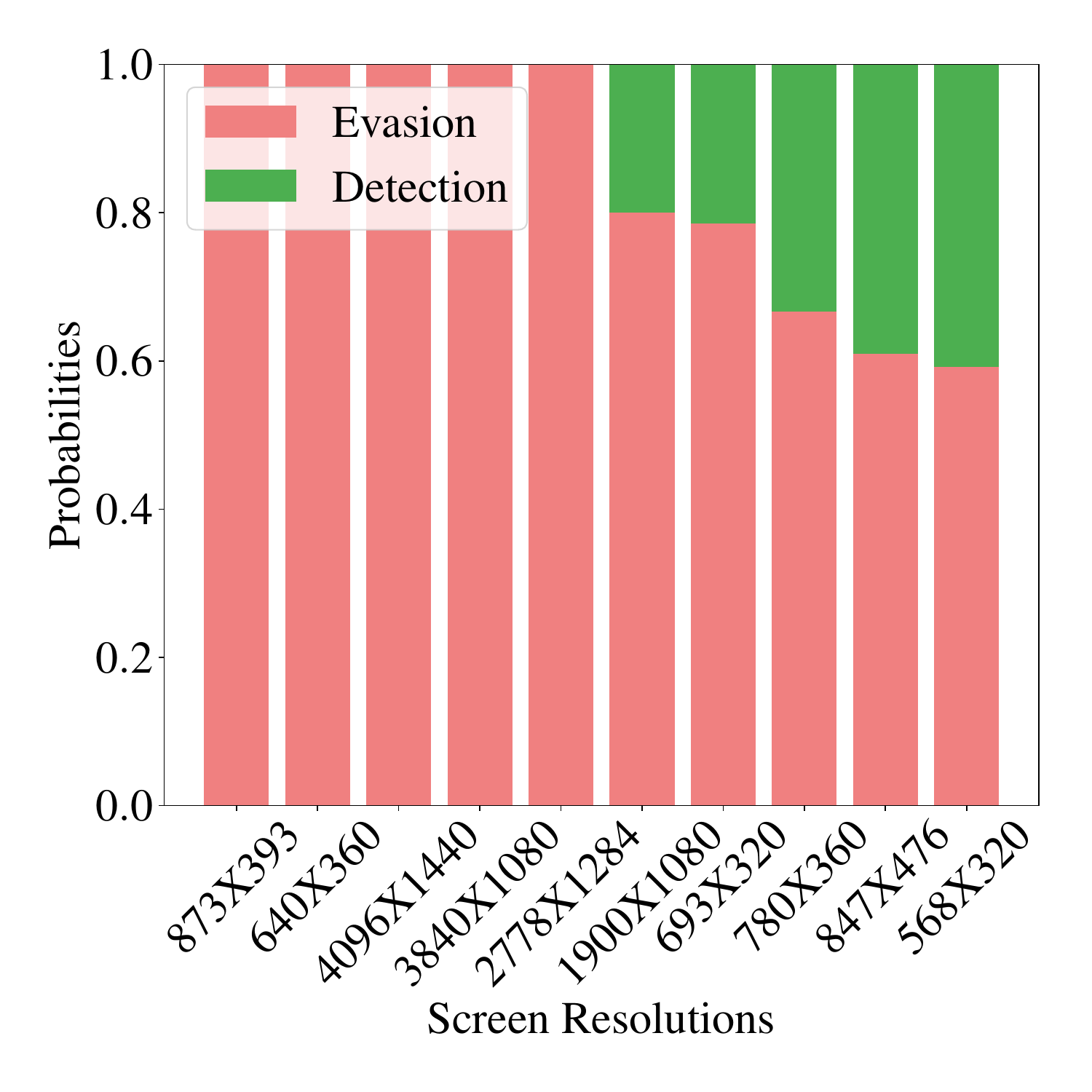}
  \caption{Bar plot showing the top 10 screen resolutions among requests received from iPhones (inferred using the User-Agent)
  that have the highest probability of evasion against \datadome. 9 out of these 10 resolutions do not exist in the real world indicating an inconsistency that can be leveraged to detect bots.}
  \label{fig:inverse_iphone_screen_res}
\end{figure}

Figure~\ref{fig:inverse_device_type} shows the top 4 device types (inferred using the \texttt{User-Agent} property 
of the browser's \texttt{navigator} object) that have the highest probability of evading \datadome among the 
requests recorded on our honey site. From the figure, we see that iPhones have the highest probability of 
evasion (around 50\%). We now look at other fingerprint attributes to determine if \smartbots sent requests 
from real iPhones or if they altered the \texttt{navigator} object on their browser to have their devices 
appear as iPhones. Since iPhones have a fixed set of screen resolutions (12 resolutions~\cite{iphone_screen_res}), 
we inspect the spread of screen resolutions captured on requests from iPhones. Upon inspection, we found 83 unique 
screen resolutions from iPhones, out of which 42 were present among those requests from iPhones that evaded 
\datadome. We also find that 9 out of the top 10 screen resolutions that have the highest probability of evading 
\datadome among requests claiming to use iPhones do not exist in the real world. We visualize these probabilities 
in Figure~\ref{fig:inverse_iphone_screen_res}. This provides strong evidence that bots alter browser APIs to show
that they use iPhones rather than using actual iPhones.

From this evidence, we see that while bots alter browser APIs, it is difficult for them to ensure that all fingerprint 
attributes remain consistent with their alterations. Thus, inconsistencies across fingerprint attributes can be leveraged
for bot detection since real users are unlikely to have such inconsistencies. In Section \ref{sec:technique} we discuss 
our systematic, data-driven, semi-automatic approach to discover such inconsistencies to improve bot detection.
\newline\newline
\fbox{\begin{minipage}{0.96\columnwidth}\textbf{Takeaway 4:} While bots alter fingerprint attributes for evasion, 
they do not ensure that all attributes are consistent with their alteration.  A particular value for a given attribute 
mapping to a large number of values for another attribute provides an avenue to discover inconsistencies.
\end{minipage}
\label{takeaway:spatial}
}

\begin{figure}
    \centering
    \includegraphics[width=\columnwidth]{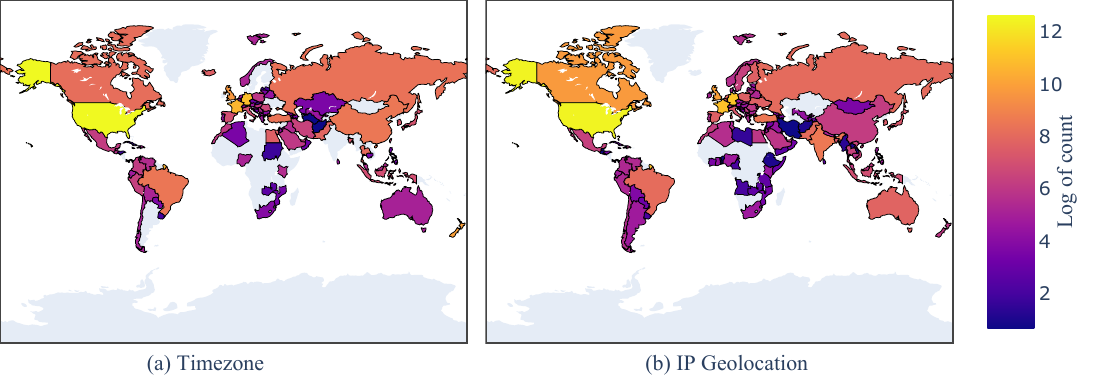}
    \caption{Plots showing a heatmap of the geographical location of requests
            inferred using the timezone attribute of the navigator object and the IP
            address. Different regions lighting up in the two heatmaps indicate that
            while bots alter the navigator object or IP address or both to change their
            geographical location, they do not ensure that the location inferred using
            both is consistent.}
    \label{fig:geographical}
\end{figure}

\subsection{Inconsistencies across fingerprint attributes and IP addresses}
\label{sec:geo_spatial}
Some \botservs advertised sending traffic from specific geographic regions (USA, Mexico,
France, etc). Having this ability to send requests from specific regions suggests that the 
\botservs are likely altering attributes that capture the geographical location of their devices.
This alteration introduces potential inconsistencies if the \botservs did not ensure that all 
attributes point to the same region.

We analyzed requests from 4 different \botservs who claimed to send requests from the 
United States, Canada, Europe, and France respectively. We first used MaxMind's GeoLite2 
database \cite{maxmind-geoip} to extract the geolocation from the IP address of the requests
from these services. We took a conservative approach when determining if the inferred geolocation
matched the region advertised by the \botserv. Concretely, we considered locations at the same UTC
offset to be a match. For example, when analyzing requests from the \botserv who advertised sending 
requests from France, we considered all requests whose geolocations mapped to any valid UTC offset that 
could overlap with France (such as Europe/Berlin) to also originate from France. With this approach, 
over 90\% of requests from each of the 4 \botservs matched the advertised geographical 
location.

However, we observed significant differences when repeating the same analysis using the browser’s 
timezone API~\cite{get-timezone-offset} to infer location. We still used the same conservative 
approach and merely replaced the geolocation inferred from the IP address with the timezone. 
Only 76.52\% of requests mapped to UTC offsets in Canada among the requests 
from the \botserv that advertised traffic from Canada. More alarmingly, we observed that only 
56\% of requests mapped to UTC offsets in Europe among the requests from the \botserv that 
advertised traffic from Europe. In contrast, we observed 92.44\% of requests to originate 
from Canada and 99.83\% of requests to originate from Europe from the corresponding 
\botservs when inferring the geolocation from the IP address. Motivated by these results, 
we visualize the geographical spread of requests based on both approaches in Figure~\ref{fig:geographical}. 
The figure reveals a number of inconsistencies in geographical locations which also constitute
inconsistencies for bot detection.
\newline\newline
\fbox{\begin{minipage}{0.96\columnwidth}\textbf{Takeaway 5:} Bots alter their IP addresses,
fingerprint attributes or both to fulfill promises of sending requests from specific locations.
\end{minipage}
\label{takeaway:temporal_inconsistency}
}

\begin{figure}[h]
\centering
  \includegraphics[width=\columnwidth]{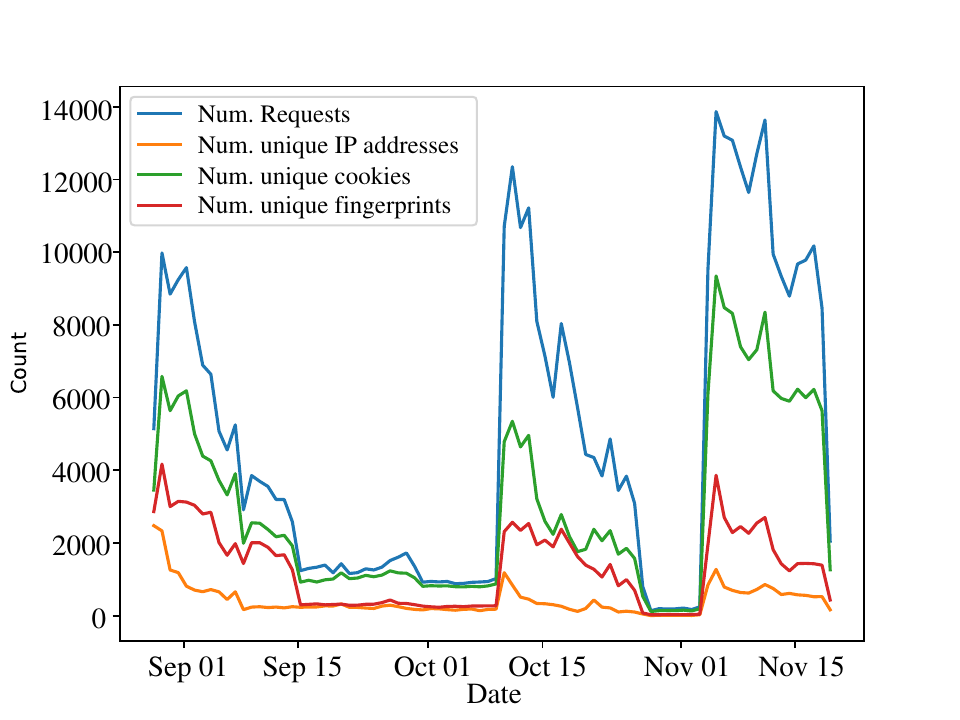}
  \caption{Temporal distribution of traffic on our honey site.}
  \label{fig:temporal}
\end{figure}
\subsection{Inconsistencies across time}
\label{sec:temporal}
Figure \ref{fig:temporal} shows the temporal spread of requests received on our honey site over 
time. The plot shows the number of requests, the number of unique IP addresses, the number of unique 
values for Cookies set by our honey site, and the number of unique FingerprintJS fingerprints 
seen per day. 

From the figure, we see that even after 2 months, we receive requests with previously unseen 
fingerprints and IP addresses. More interestingly, the spikes in the plot correspond to the days when 
we renewed our purchases. These spikes indicate that the \botservs could have access to a 
large number of devices with different device configurations that result in different browser 
attributes, and thus, different fingerprints. However, we suspect that they have a fixed set 
of devices but alter fingerprint attributes to create the illusion of sending requests from a large 
number of devices.
\begin{figure}[t]
\centering
  \includegraphics[width=\columnwidth]{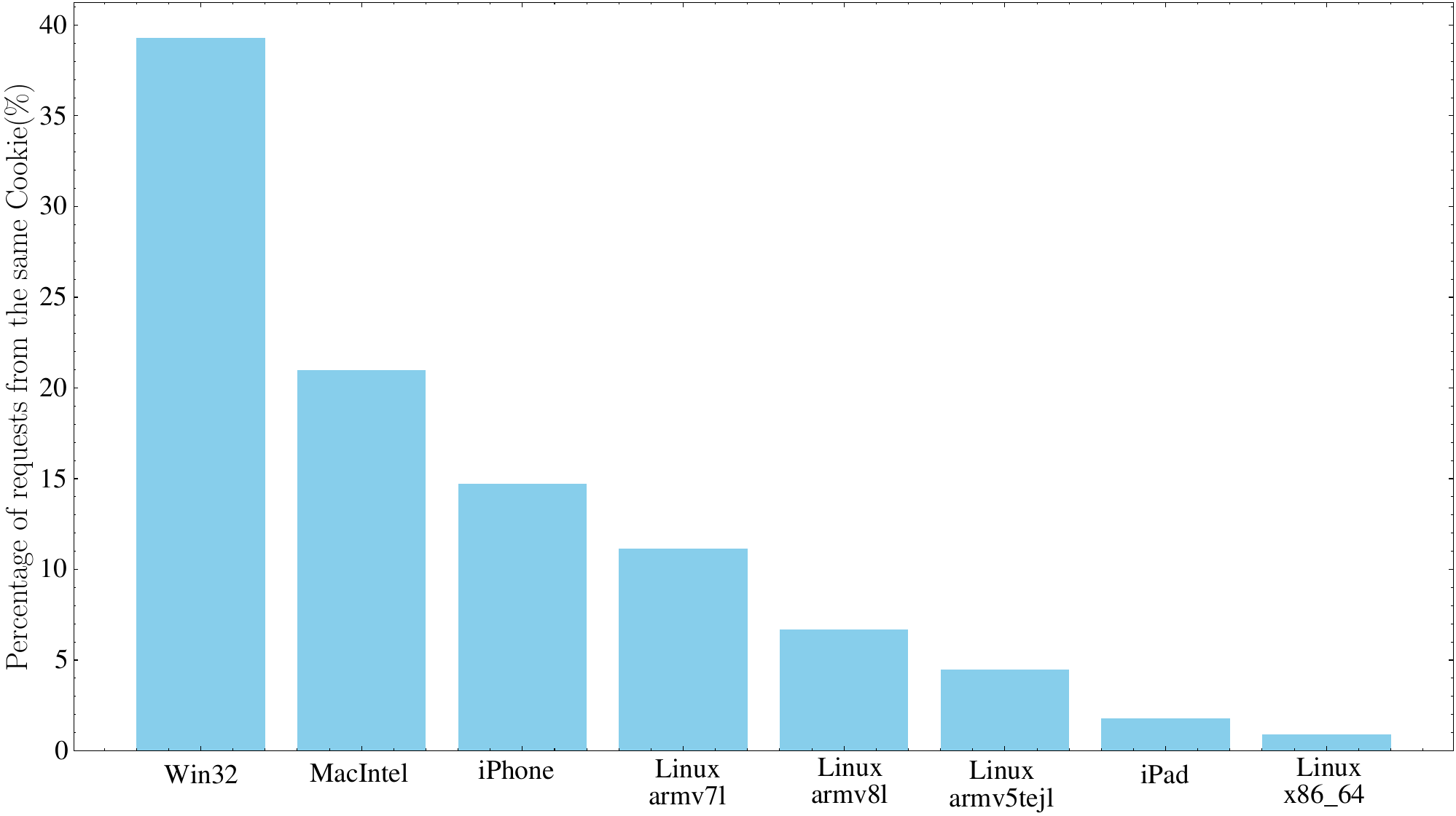}
  \caption{Percentage of requests seen across different values of the \texttt{platform}
  attribute of the \texttt{navigator} object for the same Cookie (same device). The diverse 
  spread of values provides strong evidence of bots altering the \texttt{platform} attribute 
  since it cannot change otherwise for the same device.}
  \label{fig:platform_temporal}
\end{figure}

To provide evidence that bots alter their fingerprint attributes, we inspect
the \texttt{navigator} object's \texttt{platform} attribute on all requests that
share the most commonly seen Cookie. Whenever a device sends a request to our honey 
site, we store a large random number in a first-party Cookie if it had not been set previously. 
Thus, requests bearing the same value for this Cookie should originate from the same device. Since 
the \texttt{platform} property of the \texttt{navigator} object captures information about the  
type of processor on a given device, it can never change for that device unless the entity controlling 
the device has intentionally altered the attribute. In Figure \ref{fig:platform_temporal}, we see a 
wide distribution for the navigator's platform property for the device identified as sending us the 
largest number of requests with the same Cookie. Differing values for fingerprint attributes that 
cannot change for a given device constitutes a temporal inconsistency that can be used for bot detection.
\newline\newline
\fbox{\begin{minipage}{0.96\columnwidth}\textbf{Takeaway 6:} Bots alter fingerprint
attributes to create an illusion of sending requests from a large number 
of devices. Recording differing values for fingerprint attributes that cannot
change for a given device also constitute inconsistencies to detect bots.
\end{minipage}
\label{takeaway:other_temporal_inconsistency}
}
\section{\system}
\label{sec:technique}
Our measurements in Section \ref{sec:inconsistency_measurement} reveal that there exist inconsistencies in different fingerprint attributes for a given request as well as multiple requests from the same device at
different points in time.
In this section, we present our approach to use these inconsistencies to enhance bot detection.
%
%
We categorize inconsistencies into two types: spatial and temporal.

\textbf{Spatial inconsistencies} refer to attribute values within a request that conflict or are incompatible with other attribute values in that same request.
Examples include differing locations inferred from an IP address and time zone, or implausible combinations, such as an iPhone without touch input support.
Our takeaways in Section~\ref{sec:other_spatial} and Section ~\ref{sec:geo_spatial}
show that \smartbots incur significant spatial inconsistencies across information captured in their fingerprint attributes as well as IP addresses.

\textbf{Temporal inconsistencies} are attribute values that are incompatible across different requests from the same user or users.
Examples include significantly different time zones for requests from the same IP address and inconsistent device memory values for the same Cookie value.
Our takeaway from Section \ref{sec:temporal} shows that \smartbots give rise to significant temporal inconsistencies by changing their attributes.

\subsection{Identifying spatial inconsistencies}
Our methodology for detecting spatial inconsistencies relies on the understanding that real devices can only possess a limited number of hardware and software configurations.
In contrast, bots, in their attempts to mimic real devices and evade detection, as described in Section \ref{sec:temporal}, often modify these configurations. 
However, these alterations typically do not account for every possible source of device information (such as JavaScript APIs, User-Agent, etc.), leading to a proliferation of device configurations.
This is especially noticeable in devices such as iPhones or iPads that are commonly owned by real users and have the highest success rate in evading detection (as shown in Section \ref{sec:other_spatial}).
Consequently, the increased number of bots pretending to be popular devices results in a greater variety of configurations in the dataset of requests obtained on our honey site.

However, identifying such inconsistencies is challenging because analyzing all possible attribute combinations is infeasible.
To facilitate the analysis, we first categorize attributes into different groups based on the type of information each attribute provides. 
For instance, attributes like \texttt{Color Depth}, \texttt{Screen Resolution}, and \texttt{Touch Support} are grouped because they all convey information about the device’s screen.
Table \ref{tab:features_by_category} in Appendix \ref{app:inconsistency_groups} shows the various groups used in our analysis, demonstrating how we categorize attributes to streamline the detection of inconsistencies.
\begin{figure}[!htpb]
    \centering
    \includegraphics[width=0.75\columnwidth]{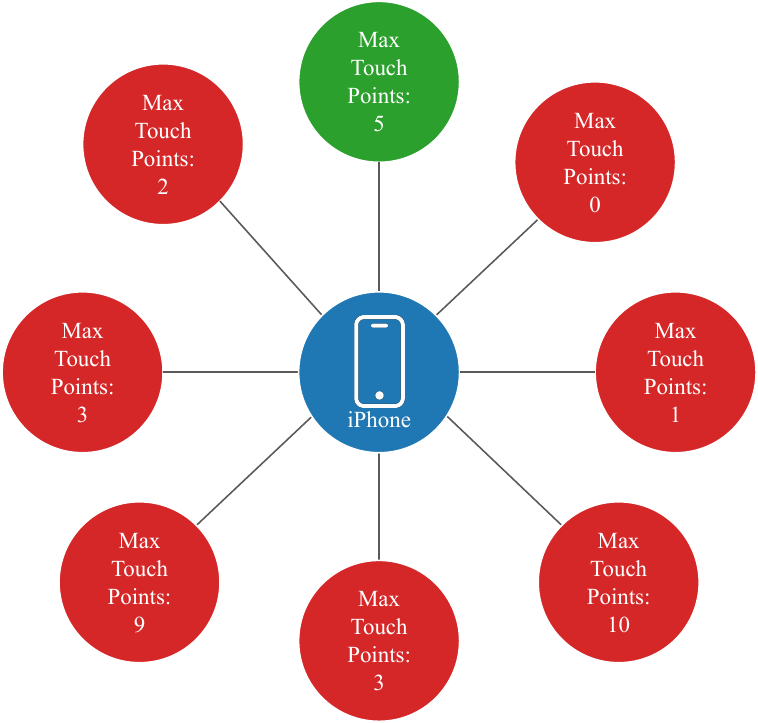}
    \caption{An example of excessive configurations of a device (iPhone) with the fingerprint attribute representing maximum touch points.}
    \label{fig:network_graph_example}
\end{figure}

Next, we analyze pairs of attributes within each category to identify spatial inconsistencies. 
For each pair, we rank the attributes based on the number of unique instances recorded in our dataset. 
For example, in the pair \texttt{UA Device} and \texttt{Maximum Touch Points}, we sort \texttt{UA Device} in descending order by the number of unique \texttt{Max Touch Points} values associated with it.
A genuine iPhone can only have five simultaneous touch points.
However, when bots imitate iPhones but report a different number of touch points, our dataset reveals an implausible number of unique combinations between \texttt{UA Device} and \texttt{Max Touch Points}.
We start with the UA Device instance that has the highest number of unique combinations and identify cases where the combination of these two attributes is impossible.
After identifying the inconsistent pair of attribute values, we repeat the process with lower-ranked unique combinations and other attribute pairs.
Appendix \ref{app:alg} defines our algorithm to identify spatial inconsistencies.
This algorithm helps us identify the most frequently altered attributes and the spatial inconsistencies they produce.
Table \ref{table:inconsistencies_identified} in Appendix \ref{app:inconsistencies_identified} provides examples of such inconsistencies in our dataset.

\subsection{Identifying temporal inconsistencies}
Building upon our findings in Section \ref{sec:temporal}, we utilize both the large random number identifier set by our honey sites in each visiting device's browser storage (Cookie) and IP address to identify temporal inconsistencies.
First, we use the Cookie identifier to measure variance in immutable device attributes (e.g., number of CPU cores, device memory) across requests containing the same identifier.
If an incoming request increases the number of unique attribute values associated with previous identifiers, we consider that request to be temporally inconsistent.
For instance, if all previous requests from a device have a \texttt{Hardware Concurrency} value of 4 and a new request contains a value of 6, we label that request as temporally inconsistent.

We also use a user's IP address to identify temporal inconsistencies related to time zones and location.
If an incoming request increases the number of unique time zones (measured as an offset from UTC) associated with that IP, we classify that request as temporally inconsistent.
Similarly, we also identify temporal inconsistencies in location information provided through the IP address and \texttt{navigator.geolocation}.

\begin{table*}[t]
    \centering
    \caption{Improvement in \datadome and \botd's detection rate on traffic from each \botserv when incorporating \system.}
        \small
          \resizebox*{0.9\textwidth}{!}{%
            \begin{tabular}{l c c c c c}
            \toprule
        {\bf{Bot}} & {\bf{Num.}} & {\bf{DataDome}} & {\bf{DataDome + \system}} & {\bf{BotD}} & {\bf{BotD + \system}}\\
        {\bf{Service}} & {\bf{Requests}} &  {\bf{Detection  Rate}} & {\bf{Detection Rate}} & {\bf{Detection Rate}} & {\bf{Detection Rate}}\\
        \midrule
        {S1} & {121500} & {55.99\%} & {83.41\%} & {28.42\%} & {60.26\%}\\
        {S2} & {63708} & {57.01\%} & {82.61\%} & {27.71\%} & {55.83\%}\\
        {S3} & {54746} & {25.09\%} & {46.31\%} & {89.74\%} & {94.17\%}\\ 
        {S4} & {47278} & {61.35\%} & {82.35\%} & {26.15\%} & {52.09\%}\\
        {S5} & {40087} & {76.14\%} & {88.19\%} & {27.35\%} & {50.46\%}\\
        {S6} & {32447} & {28.19\%} & {43.7\%} & {94.55\%} & {97.05\%}\\
        {S7} & {28940} & {97.44\%} & {99.35\%} & {360.01\%} & {83.91\%}\\
        {S8} & {26335} & {19.57\%} & {47.84\%} & {71.1\%} & {86.06\%}\\
        {S9} & {23412} & {27.71\%} & {65.69\%} & {80.67\%} & {94.07\%}\\
        {S10} & {18967} & {84.23\%} & {94.7\%} & {40.64\%} & {70.43\%}\\
        {S11} & {17996} & {93.45\%} & {98.63\%} & {59.36\%} & {80.16\%}\\
        {S12} & {7010} & {94.95\%} & {98.36\%} & {48.56\%} & {78.21\%}\\
        {S13} & {5119} & {93.04\%} & {99.1\%} & {49.48\%} & {87.04\%}\\
        {S14} & {4920} & {16.26\%} & {66.27\%} & {9.92\%} & {67.29\%}\\
        {S15} & {4219} & {88.86\%} & {99.6\%} & {0\%} & {77.87\%}\\
        {S16} & {4174} & {95.52\%} & {99.69\%} & {99.98\%} & {100\%}\\
        {S17} & {2999} & {25.34\%} & {43.88\%} & {92.1\%} & {95.1\%}\\
        {S18} & {1430} & {79.3\%} & {99.86\%} & {0\%} & {83.57\%}\\
        {S19} & {1411} & {90.08\%} & {99.5\%} & {0\%} & {59.76\%}\\
        {S20} & {382} & {2.88\%} & {7.59\%} & {2.88\%} & {7.07\%}\\
    \bottomrule
     \end{tabular}
     }
     \label{tab:service_improvement}
     \end{table*}
\subsection{Improved bot detection}
\label{subsec:accuracy_improvement_filterlist}
In this section, we describe our methodology to use temporal and spatial inconsistencies to detect bots that evade \datadome and \botd.
To measure the improvement in accuracy from spatial inconsistencies, we translate the inconsistencies identified in Table \ref{table:inconsistencies_identified} into filter rules.
These filter rules are then matched with each request that evaded detection by \datadome or \botd.
For temporal inconsistencies, we use the timestamp of each request to determine the order in which requests were made, applying filter rules to identify inconsistencies created by requests arriving later.

The results in Table \ref{table:comparison_accuracy_improvement} show that using rules generated through spatial and temporal inconsistency analysis can decrease the evasion of bots against \botd by \botdevasiondec and against \datadome by \datadomeevasiondec. Table \ref{tab:service_improvement} shows the improvement in detection on requests obtained from each individual \botserv. 
We evaluated the generalizability of our methodology by computing filter rules on 80\% of the requests obtained on our honey site and evaluating them on the remaining 20\%. 
This evaluation led to a meagre drop in detection accuracy of 0.42\% for \botd and 0.23\% for \datadome, thereby showing that \system generalizes to unseen requests.

%

\begin{table}[ht]
\centering
\caption{Comparison of the improvement in \datadome and \botd's detection accuracies 
resulting from different forms of inconsistency analysis.}
\begin{tabular}{l c c}
\toprule
\textbf{} & \textbf{\datadome} & \textbf{\botd} \\
\midrule
\textbf{None}                   &             55.44\% &                47.07\%  \\
\textbf{Spatial}                &               76.04\% &                 70.33\% \\
\textbf{Temporal}               &               56.53\% &                 48.09\% \\
\textbf{Combined}               &               76.88\% &                 70.86\% \\
\bottomrule
\end{tabular}
\label{table:comparison_accuracy_improvement}
\end{table}

Our results on the requests received on our honey site show that using a filter list to counter commonly found inconsistencies is an effective method to detect and block \smartbots.
Filter lists are commonplace in the anti-tracking community, where they provide a good trade-off between performance and accuracy in detecting advertising and tracking services.
Currently, no such alternative exists to detect bots that show inconsistent fingerprints.
Our methodology is a first step towards creating such filter lists to enhance online bot detection.

\subsection{Real user traffic}
\label{sec:real_user_eval}
We also evaluate \system's filter rules against traffic from real users to ensure that 
our improvements in bot detection do not incorrectly detect real users as bots.
Concretely, we shared a version of our honey site that contained a unique URL with students at our university. Since we only shared this URL with bonafide
students, we have high confidence that requests from real users were recorded at this URL. We did not 
collect any Personally Identifiable Information (PII) from these users and discuss the ethics of collecting this data in Appendix \ref{app:ethics}.

We report a true negative rate of 96.84\% on the 2,206 requests received at this URL. 
The small number of 
false positives were likely due to students experimenting with User-Agent spoofers, as these cases triggered 
spatial inconsistencies involving User-Agents. 
We could not conduct large-scale evaluation on real user traffic 
in the wild since it would be challenging to ensure ground-truth. Regardless, our evaluation shows low false 
positive rates, which can be further mitigated using CAPTCHAs if needed (Section \ref{sec:fp_disc}).

\subsection{Privacy-enhancing browsers}
\label{sec:pet_eval}
Privacy-enhancing browsers such as Brave \cite{brave-browser},  Tor \cite{tor-browser}, and
Fingerprint Spoofer \cite{fingerprint-spoofer} alter fingerprint attributes to protect 
user privacy against tracking \cite{fp-inspector, oneclass-tracker-detection}.
In this section, we examine the attributes altered by such technologies and their impact 
on \system.

We conducted an experiment where we sent requests to different versions of our honey site (each with a 
distinct URL) while employing five different privacy-enhancing browsers: Safari, Brave, Tor browsers 
as well as uBlock Origin and AdBlockPlus browser extensions on Google Chrome.  We 300 requests from devices 
running macOS (M1 MacBook Pro), Linux (Intel Coffee Lake Desktop), iOS (iPad Pro), and Android (Google Pixel 7).

\textbf{Brave}
Brave browser currently alters 6 different fingerprint attributes: \texttt{audio}, \texttt{canvas}, \texttt{plugins}, \texttt{deviceMemory}, \texttt{hardwareConcurrency}, and
\texttt{screenResolution}. Our inconsistency rules do not
incorporate the former three attributes and Brave's alterations to the 
others were consistent with other attributes. For instance, Brave alters \texttt{deviceMemory} on desktops to plausible 
values (0.5, 1, 2, 4, and 8), which align with the amount of memory in typical desktops and remain 
consistent with other fingerprint attributes.

However, since Brave browser retains Cookies across requests, the requests triggered 
several temporal inconsistencies where multiple requests shared the same Cookie but had differing 
values for both \texttt{hardwareConcurrency} and \texttt{deviceMemory}. Such inconsistencies are 
rare in real-world scenarios, as they require users to enable Brave’s fingerprint protection while 
retaining Cookies. These rare false positives can be mitigated using CAPTCHAs, with the verification 
result stored in Cookies (Section \ref{sec:fp_disc}).

Although \system does not detect requests from Brave browser as bots, we argue that \botservs cannot exploit Brave for 
evasion since they seek to alter attributes that are not supported by Brave.  Concretely, we see that Brave only 
alters 2 attributes that are most commonly altered by \smartbots (Section \ref{sec:measurement_attributes}) and
does not alter other attributes that are of interest to bots such as those pertaining to their device type or 
geolocation (Section \ref{sec:inconsistency_measurement}). If Brave were to alter more fingerprint attributes 
in the future, \system could become prone to more false positives. However, even in such a hypothetical scenario, 
these false positives can be mitigated using CAPTCHAs (Section \ref{sec:fp_disc}. Moreover, only a small set of users would 
encounter these CAPTCHAs if Brave's market share continues to remain at 1\% \cite{brave-market-share}.

\textbf{Tor}
\system detected all requests from Tor browser as bots since they triggered spatial inconsistencies between 
the geolocation inferred from their IP address and the \texttt{timezone} attribute of their navigator object.
While Tor results in false positives, we expect a small set of users to be affected since Tor likely has 
less than 1\% market share \cite{browser_market}. Furthermore, most websites currently block requests from Tor 
\cite{tor-block}, due to the difficulty in distinguishing Tor traffic from bots. To mitigate false positives, 
we can present users with CAPTCHAs rather than blocking their requests (Section \ref{sec:fp_disc}). We
report the detection accuracy of \datadome and \botd on Brave and Tor traffic in Appendix \ref{app:tor_brave_dd_botd}.

\textbf{Safari, uBlock Origin, and AdBlockPlus}
None of these requests were detected as bots. This is because these tools 
protect privacy by blocking tracking requests rather than altering fingerprint attributes.
The two extensions cater to over 80 million users combined on the most widely used browser, 
Google Chrome \cite{browser_market,ublock_webstore,adblock_webstore}. Safari has the 
second largest market share among web browsers \cite{browser_market}. This shows that 
\system can detect bots while having zero impact of all these users.
\section{Discussion}
\subsection{Overcoming false positives}
\label{sec:fp_disc}
Our evaluation on requests from real users show that \system incurs low, but 
non-zero false positive rates (Section \ref{sec:real_user_eval}). Our experiments with
privacy-enhancing technologies also reveal certain scenarios that could lead to false
positives (Section \ref{sec:pet_eval}). In the context of this paper, false positives
refer to requests from real users that were incorrectly detected as bots. 
Challenging users to solve CAPTCHAs rather than blocking them offers a promising 
solution to mitigate false positives \cite{webrunner-2049, boxer}. While effective,
CAPTCHAs could potentially frustrate certain users \cite{cacti, captcha_annoyance}.
This frustration can be mitigated by storing the result of a
CAPTCHA verification in a Cookie, thereby reducing the frequency at which users are asked to solve CAPTCHAs.

\subsection{Improving \system}
Inconsistencies provide a promising avenue for detection as long as there exist at least one pair of attributes that cannot exist in the real world.
Accordingly, increasing the number of captured attributes introduces more opportunities for inconsistencies which can be leveraged for detection.
In this paper, we confined \system to only look for inconsistencies among HTTP headers and the attributes captured by FingerprintJS.
Incorporating other attributes such as those from CreepJS \cite{creepjs} can further improve \system.
%
%
%

Researchers have proposed side-channels based on physical device characteristics to uniquely identify devices even among those with identical hardware and software configurations \cite{clock-skew, drawn_apart-gpu, centauri, rowhammer-puf}.
Such techniques can significantly empower temporal inconsistencies to detect bots. With \system, we used Cookies to identify requests that originated from the same device.
Bots will be able to overcome our temporal inconsistencies by merely deleting their cookies. Bots would not be able to  drop unique identifiers that originate from the physical properties of hardware that cannot 
be modified.
However, capturing more attributes as well as capturing persistent identifiers pose threats to privacy.

\subsection{Deployment of filter list rules}
\system generates filter lists of inconsistencies to improve bot detection (Section \ref{subsec:accuracy_improvement_filterlist}).
The anti-tracking community \cite{webgraph, adgraph, cookiegraph, AdblockPlus, AdguardFilters, uBlock} typically incorporates
filter lists on the client side using browser extensions to block the execution of tracking requests and other resources.
Similarly, we envision \antibotservs such as \datadome and \botd to include \system's filter lists as part of their client-side
scripts improved for bot detection.

\subsection{Limitations}
Our results show that \system's rules improve the detection of \smartbots.
\Smartbots will be able to overcome \system if they evolve to ensure that they can alter fingerprint attributes without introducing any inconsistencies.
Incorporating unmodifiable attributes provides a robust solution to enhance \system, but such attributes also pose threats to
privacy.
\subsection{Coexistence of Bot Detection and Privacy-Enhancing Technologies}
Our evaluation (Section \ref{sec:pet_eval}) shows that \system can detect bots without impeding most commonly used privacy-enhancing 
technologies (except Tor). This observation is interesting because many assume that the goals of bot detection and online tracking are 
identical. They believe that enhancements to bot detection would also bolster online tracking and weaken privacy-enhancing tools.

While bot detection and tracking overlap, tracking is more complex as it seeks to uniquely identify each user. In contrast, 
bot detection solely seeks to determine if a particular request was generated by a bot. Accordingly,
altering any fingerprint attribute enhances privacy by making it harder for trackers to link requests from the same user, 
even when other attributes remain unchanged. On the other hand, altering any fingerprint attribute does not necessarily 
help bots with evasion since other attributes can still reveal their presence. This distinction between bot detection
and online tracking allows bot detection systems like \system to coexist with privacy-enhancing technologies.

However, given the overlap, certain enhancements to bot detection such as incorporating more attributes or incorporating
unmodifiable attributes can threaten user privacy. Future research focusing on
privacy-preserving bot detection such as identifying the intent behind trackers to not block those indulging in bot 
detection or an in-browser detection mechanism can bridge the gap to potentially address concerns of privacy protection 
as well as bot detection.
\section{Conclusion}
We find evidence that bots alter fingerprint attributes to evade detection. 
However, we find evidence that such \smart bots end up introducing inconsistencies among the fingerprint attributes that can be used for more reliable bot detection. 
We propose \system, a data-driven, semi-automatic approach to discover inconsistencies in fingerprint attributes for detecting \smart bots in the wild that are able to evade detection by \antibotservs. 
As the arms race between \smart bots and \antibotservs evolves, it remains to be seen whether bots can alter their fingerprint attributes while avoiding inconsistency. 
We believe that it would be challenging for bots to do so because a browser fingerprint is a high dimensional feature set with numerous --  often subtle -- correlations between attributes that are difficult to anticipate and account for when altering fingerprints. 
Put simply, it is challenging to tell a complex lie while keeping the story always straight. 
While \system rule generation approach may need to be evolved to generate rules for other types of consistencies for future generation of bots, we believe the basic principle will stand over time. 


\bibliographystyle{ACM-Reference-Format}
\bibliography{refs}


\begin{thebibliography}{69}


\ifx \showCODEN    \undefined \def \showCODEN     #1{\unskip}     \fi
\ifx \showDOI      \undefined \def \showDOI       #1{#1}\fi
\ifx \showISBNx    \undefined \def \showISBNx     #1{\unskip}     \fi
\ifx \showISBNxiii \undefined \def \showISBNxiii  #1{\unskip}     \fi
\ifx \showISSN     \undefined \def \showISSN      #1{\unskip}     \fi
\ifx \showLCCN     \undefined \def \showLCCN      #1{\unskip}     \fi
\ifx \shownote     \undefined \def \shownote      #1{#1}          \fi
\ifx \showarticletitle \undefined \def \showarticletitle #1{#1}   \fi
\ifx \showURL      \undefined \def \showURL       {\relax}        \fi
\providecommand\bibfield[2]{#2}
\providecommand\bibinfo[2]{#2}
\providecommand\natexlab[1]{#1}
\providecommand\showeprint[2][]{arXiv:#2}

\bibitem[{abrahamjuliot}({[n.\,d.]})]%
        {creepjs}
\bibfield{author}{\bibinfo{person}{{abrahamjuliot}}.}
  \bibinfo{year}{[n.\,d.]}\natexlab{}.
\newblock \bibinfo{title}{{CreepJS}}.
\newblock
  \bibinfo{howpublished}{\url{https://github.com/abrahamjuliot/creepjs}}.
\newblock


\bibitem[AdguardTeam({[n.\,d.]})]%
        {AdguardFilters}
\bibfield{author}{\bibinfo{person}{AdguardTeam}.}
  \bibinfo{year}{[n.\,d.]}\natexlab{}.
\newblock \bibinfo{title}{Adguard Filters}.
\newblock
  \bibinfo{howpublished}{\url{https://github.com/AdguardTeam/AdguardFilters}}.
\newblock


\bibitem[Amin~Azad et~al\mbox{.}(2020)]%
        {webrunner-2049}
\bibfield{author}{\bibinfo{person}{Babak Amin~Azad}, \bibinfo{person}{Oleksii
  Starov}, \bibinfo{person}{Pierre Laperdrix}, {and} \bibinfo{person}{Nick
  Nikiforakis}.} \bibinfo{year}{2020}\natexlab{}.
\newblock \showarticletitle{{Web Runner 2049: Evaluating Third-Party Anti-bot
  Services}}. In \bibinfo{booktitle}{\emph{{DIMVA 2020 - 17th Conference on
  Detection of Intrusions and Malware \& Vulnerability Assessment}}}.
  \bibinfo{address}{Lisboa / Virtual, Portugal}.
\newblock
\urldef\tempurl%
\url{https://hal.science/hal-02612454}
\showURL{%
\tempurl}


\bibitem[Askari et~al\mbox{.}(2024)]%
        {twitterbot}
\bibfield{author}{\bibinfo{person}{Hadi Askari}, \bibinfo{person}{Anshuman
  Chhabra}, \bibinfo{person}{Bernhard~Clemm von Hohenberg},
  \bibinfo{person}{Michael Heseltine}, {and} \bibinfo{person}{Magdalena
  Wojcieszak}.} \bibinfo{year}{2024}\natexlab{}.
\newblock \bibinfo{title}{Incentivizing News Consumption on Social Media
  Platforms Using Large Language Models and Realistic Bot Accounts}.
\newblock
\newblock
\showeprint[arxiv]{2403.13362}~[cs.SI]


\bibitem[Asuman~Senol and Bilogrevic(2024)]%
        {double-edged-sword}
\bibfield{author}{\bibinfo{person}{Dylan~Cutler Asuman~Senol, Alisha~Ukani}
  {and} \bibinfo{person}{Igor Bilogrevic}.} \bibinfo{year}{2024}\natexlab{}.
\newblock \showarticletitle{The Double Edged Sword: Identifying Authentication
  Pages and their Fingerprinting Behavior}.
\newblock


\bibitem[{Babylon Traffic}({[n.\,d.]})]%
        {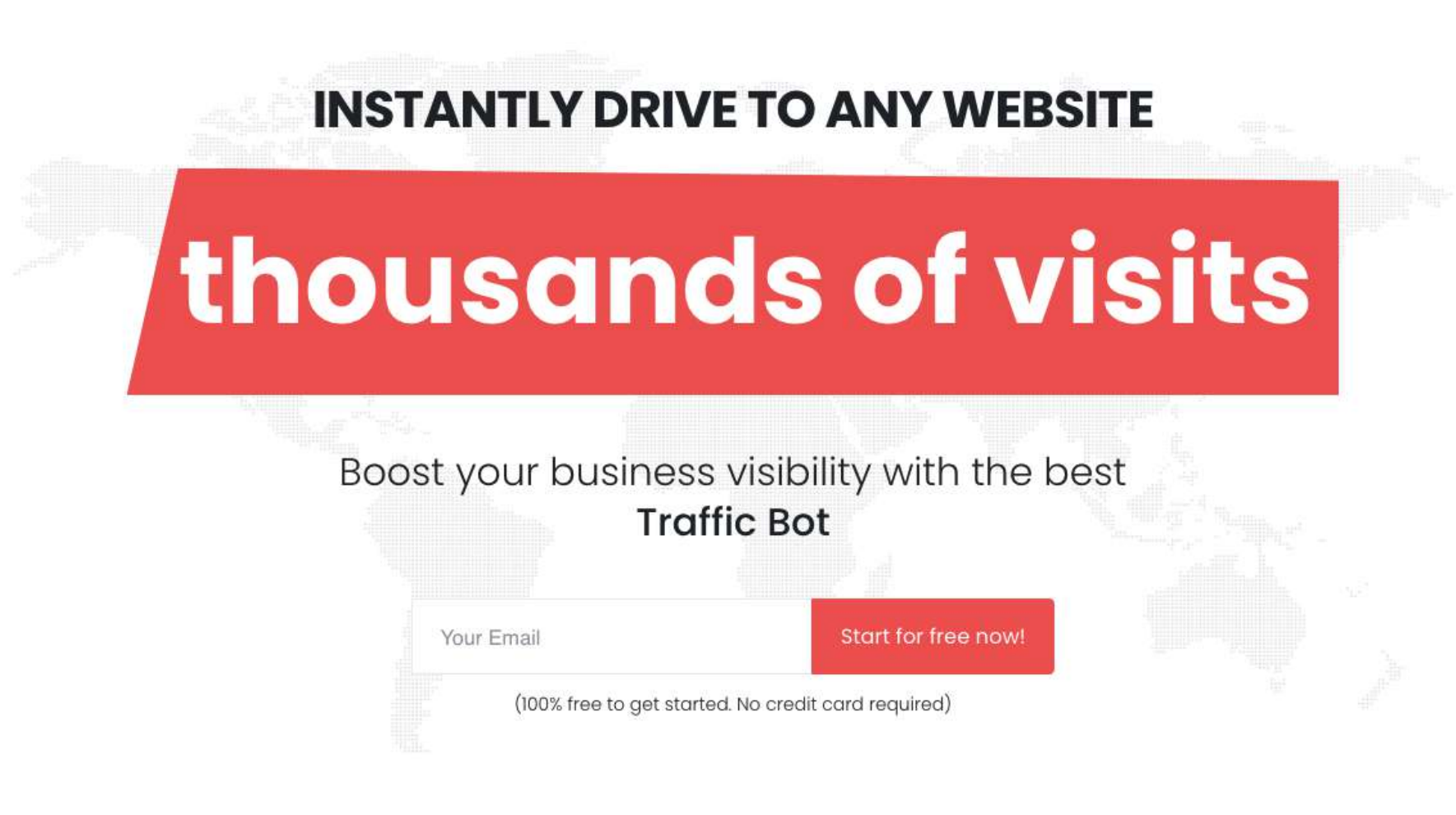}
\bibfield{author}{\bibinfo{person}{{Babylon Traffic}}.}
  \bibinfo{year}{[n.\,d.]}\natexlab{}.
\newblock \bibinfo{title}{{Boost your business visibility with the best Traffic
  Bot}}.
\newblock \bibinfo{howpublished}{\url{https://www.babylontraffic.com/}}.
\newblock


\bibitem[{Brave}({[n.\,d.]})]%
        {brave-browser}
\bibfield{author}{\bibinfo{person}{{Brave}}.}
  \bibinfo{year}{[n.\,d.]}\natexlab{}.
\newblock \bibinfo{title}{{Secure, Fast, \& Private Web Browser with Adblocker
  | Brave}}.
\newblock \bibinfo{howpublished}{\url{https://brave.com/}}.
\newblock


\bibitem[{brianhama}({[n.\,d.]})]%
        {brianhama}
\bibfield{author}{\bibinfo{person}{{brianhama}}.}
  \bibinfo{year}{[n.\,d.]}\natexlab{}.
\newblock \bibinfo{title}{{bad-asn-list}}.
\newblock
  \bibinfo{howpublished}{\url{https://github.com/brianhama/bad-asn-list/tree/master}}.
\newblock


\bibitem[Cabri et~al\mbox{.}(2018)]%
        {Cabri2018HPCC}
\bibfield{author}{\bibinfo{person}{Alberto Cabri}, \bibinfo{person}{Grażyna
  Suchacka}, \bibinfo{person}{Stefano Rovetta}, {and}
  \bibinfo{person}{Francesco Masulli}.} \bibinfo{year}{2018}\natexlab{}.
\newblock \showarticletitle{Online Web Bot Detection Using a Sequential
  Classification Approach}. In \bibinfo{booktitle}{\emph{2018 IEEE 20th
  International Conference on High Performance Computing and Communications;
  IEEE 16th International Conference on Smart City; IEEE 4th International
  Conference on Data Science and Systems (HPCC/SmartCity/DSS)}}.
\newblock


\bibitem[Chen et~al\mbox{.}(2013)]%
        {fastflux}
\bibfield{author}{\bibinfo{person}{Chia-Mei Chen}, \bibinfo{person}{Sheng-Tzong
  Cheng}, {and} \bibinfo{person}{Ju-Hsien Chou}.}
  \bibinfo{year}{2013}\natexlab{}.
\newblock \showarticletitle{Detection of fast-flux domains}.
\newblock \bibinfo{journal}{\emph{Journal of Advances in Computer Networks}}
  \bibinfo{volume}{1}, \bibinfo{number}{2} (\bibinfo{year}{2013}),
  \bibinfo{pages}{148--152}.
\newblock


\bibitem[Chiapponi et~al\mbox{.}(2022)]%
        {industrial_scraping}
\bibfield{author}{\bibinfo{person}{Elisa Chiapponi}, \bibinfo{person}{Marc
  Dacier}, \bibinfo{person}{Olivier Thonnard}, \bibinfo{person}{Mohamed
  Fangar}, \bibinfo{person}{Mattias Mattsson}, {and} \bibinfo{person}{Vincent
  Rigal}.} \bibinfo{year}{2022}\natexlab{}.
\newblock \showarticletitle{An industrial perspective on web scraping
  characteristics and open issues}. In \bibinfo{booktitle}{\emph{2022 52nd
  Annual IEEE/IFIP International Conference on Dependable Systems and Networks
  - Supplemental Volume (DSN-S)}}. \bibinfo{pages}{5--8}.
\newblock
\urldef\tempurl%
\url{https://doi.org/10.1109/DSN-S54099.2022.00012}
\showDOI{\tempurl}


\bibitem[{DataDome}({[n.\,d.]})]%
        {datadome}
\bibfield{author}{\bibinfo{person}{{DataDome}}.}
  \bibinfo{year}{[n.\,d.]}\natexlab{}.
\newblock \bibinfo{title}{{Bot And Online Fraud Protection Solution}}.
\newblock \bibinfo{howpublished}{\url{https://datadome.co/}}.
\newblock


\bibitem[Dave et~al\mbox{.}(2013)]%
        {viceroi}
\bibfield{author}{\bibinfo{person}{Vacha Dave}, \bibinfo{person}{Saikat Guha},
  {and} \bibinfo{person}{Yin Zhang}.} \bibinfo{year}{2013}\natexlab{}.
\newblock \showarticletitle{ViceROI: Catching Click-Spam in Search Ad
  Networks}. In \bibinfo{booktitle}{\emph{Proceedings of the 2013 ACM SIGSAC
  Conference on Computer and Communications Security}} (Berlin, Germany)
  \emph{(\bibinfo{series}{CCS '13})}. \bibinfo{publisher}{Association for
  Computing Machinery}, \bibinfo{address}{New York, NY, USA},
  \bibinfo{pages}{765–776}.
\newblock
\showISBNx{9781450324779}
\urldef\tempurl%
\url{https://doi.org/10.1145/2508859.2516688}
\showDOI{\tempurl}


\bibitem[Din et~al\mbox{.}(2020)]%
        {boxer}
\bibfield{author}{\bibinfo{person}{Zainul~Abi Din}, \bibinfo{person}{Hari
  Venugopalan}, \bibinfo{person}{Jaime Park}, \bibinfo{person}{Andy Li},
  \bibinfo{person}{Weisu Yin}, \bibinfo{person}{HaoHui Mai},
  \bibinfo{person}{Yong~Jae Lee}, \bibinfo{person}{Steven Liu}, {and}
  \bibinfo{person}{Samuel~T. King}.} \bibinfo{year}{2020}\natexlab{}.
\newblock \showarticletitle{Boxer: Preventing fraud by scanning credit cards}.
  In \bibinfo{booktitle}{\emph{29th USENIX Security Symposium (USENIX Security
  20)}}. \bibinfo{publisher}{USENIX Association}, \bibinfo{pages}{1571--1588}.
\newblock
\showISBNx{978-1-939133-17-5}
\urldef\tempurl%
\url{https://www.usenix.org/conference/usenixsecurity20/presentation/din}
\showURL{%
\tempurl}


\bibitem[{Erez Hasson}({[n.\,d.]})]%
        {imperva_evasive}
\bibfield{author}{\bibinfo{person}{{Erez Hasson}}.}
  \bibinfo{year}{[n.\,d.]}\natexlab{}.
\newblock \bibinfo{title}{{Evasive Bots Drive Online Fraud}}.
\newblock
  \bibinfo{howpublished}{\url{https://www.imperva.com/blog/evasive-bots-drive-online-fraud-2022-imperva-bad-bot-report/}}.
\newblock


\bibitem[{Eugene Belinski}({[n.\,d.]})]%
        {iphone_screen_res}
\bibfield{author}{\bibinfo{person}{{Eugene Belinski}}.}
  \bibinfo{year}{[n.\,d.]}\natexlab{}.
\newblock \bibinfo{title}{{iOS Ref}}.
\newblock \bibinfo{howpublished}{\url{https://github.com/ebelinski/iosref}}.
\newblock


\bibitem[{F5 Inc}({[n.\,d.]})]%
        {f5_bot_defense}
\bibfield{author}{\bibinfo{person}{{F5 Inc}}.}
  \bibinfo{year}{[n.\,d.]}\natexlab{}.
\newblock \bibinfo{title}{{Bot Defense}}.
\newblock
  \bibinfo{howpublished}{\url{https://docs.cloud.f5.com/docs/how-to/advanced-security/bot-defense}}.
\newblock


\bibitem[Farooqi et~al\mbox{.}(2017)]%
        {farooqi2017characterizing}
\bibfield{author}{\bibinfo{person}{Shehroze Farooqi},
  \bibinfo{person}{Guillaume Jourjon}, \bibinfo{person}{Muhammad Ikram},
  \bibinfo{person}{Mohamed~Ali Kaafar}, \bibinfo{person}{Emiliano
  De~Cristofaro}, \bibinfo{person}{Zubair Shafiq}, \bibinfo{person}{Arik
  Friedman}, {and} \bibinfo{person}{Fareed Zaffar}.}
  \bibinfo{year}{2017}\natexlab{}.
\newblock \showarticletitle{Characterizing key stakeholders in an online
  black-hat marketplace}. In \bibinfo{booktitle}{\emph{2017 APWG Symposium on
  Electronic Crime Research (eCrime)}}. IEEE, \bibinfo{pages}{17--27}.
\newblock


\bibitem[Fingerprint({[n.\,d.]})]%
        {fingerprintjs}
\bibfield{author}{\bibinfo{person}{Fingerprint}.}
  \bibinfo{year}{[n.\,d.]}\natexlab{}.
\newblock \bibinfo{title}{FingerprintJS}.
\newblock
  \bibinfo{howpublished}{\url{https://github.com/fingerprintjs/fingerprintjs}}.
\newblock


\bibitem[{Fingerprint}({[n.\,d.]})]%
        {botd}
\bibfield{author}{\bibinfo{person}{{Fingerprint}}.}
  \bibinfo{year}{[n.\,d.]}\natexlab{}.
\newblock \bibinfo{title}{{Open-source JavaScript Bot Detection Library}}.
\newblock
  \bibinfo{howpublished}{\url{https://fingerprint.com/products/bot-detection/}}.
\newblock


\bibitem[{Fingerprint Spoofer}({[n.\,d.]})]%
        {fingerprint-spoofer}
\bibfield{author}{\bibinfo{person}{{Fingerprint Spoofer}}.}
  \bibinfo{year}{[n.\,d.]}\natexlab{}.
\newblock \bibinfo{title}{{Fingerprint Spoofer}}.
\newblock
  \bibinfo{howpublished}{\url{https://chromewebstore.google.com/detail/fingerprint-spoofer/facgnnelgcipeopfbjcajpaibhhdjgcp}}.
\newblock


\bibitem[Gianvecchio et~al\mbox{.}(2009)]%
        {Gianvecchio2009CCS}
\bibfield{author}{\bibinfo{person}{S. Gianvecchio}, \bibinfo{person}{Z. Wu},
  \bibinfo{person}{M. Xie}, {and} \bibinfo{person}{H. Wang}.}
  \bibinfo{year}{2009}\natexlab{}.
\newblock \showarticletitle{Battle of Botcraft: Fighting Bots in Online Games
  with Human Observational Proofs}. In \bibinfo{booktitle}{\emph{Proceedings of
  the 16th ACM Conference on Computer and Communications Security}}.
\newblock


\bibitem[Gianvecchio et~al\mbox{.}(2008)]%
        {Gianvecchio2008USENIX}
\bibfield{author}{\bibinfo{person}{S. Gianvecchio}, \bibinfo{person}{M. Xie},
  \bibinfo{person}{Z. Wu}, {and} \bibinfo{person}{H. Wang}.}
  \bibinfo{year}{2008}\natexlab{}.
\newblock \showarticletitle{Measurement and Classification of Humans and Bots
  in Internet Chat}. In \bibinfo{booktitle}{\emph{Proceedings of the 17th
  USENIX Symposium on Security}}.
\newblock


\bibitem[{Google}({[n.\,d.]})]%
        {google-bots}
\bibfield{author}{\bibinfo{person}{{Google}}.}
  \bibinfo{year}{[n.\,d.]}\natexlab{}.
\newblock \bibinfo{title}{{Verifying Googlebot and other Google crawlers}}.
\newblock
  \bibinfo{howpublished}{\url{https://developers.google.com/search/docs/crawling-indexing/verifying-googlebot}}.
\newblock


\bibitem[gorhill({[n.\,d.]})]%
        {uBlock}
\bibfield{author}{\bibinfo{person}{gorhill}.}
  \bibinfo{year}{[n.\,d.]}\natexlab{}.
\newblock \bibinfo{title}{uBlock Origin}.
\newblock \bibinfo{howpublished}{\url{https://github.com/gorhill/uBlock}}.
\newblock


\bibitem[Go\ss{}en et~al\mbox{.}(2021)]%
        {hlisa}
\bibfield{author}{\bibinfo{person}{Daniel Go\ss{}en}, \bibinfo{person}{Hugo
  Jonker}, \bibinfo{person}{Stefan Karsch}, \bibinfo{person}{Benjamin Krumnow},
  {and} \bibinfo{person}{David Roefs}.} \bibinfo{year}{2021}\natexlab{}.
\newblock \showarticletitle{HLISA: towards a more reliable measurement tool}.
  In \bibinfo{booktitle}{\emph{Proceedings of the 21st ACM Internet Measurement
  Conference}} (Virtual Event) \emph{(\bibinfo{series}{IMC '21})}.
  \bibinfo{publisher}{Association for Computing Machinery},
  \bibinfo{address}{New York, NY, USA}, \bibinfo{pages}{380–389}.
\newblock
\showISBNx{9781450391290}
\urldef\tempurl%
\url{https://doi.org/10.1145/3487552.3487843}
\showDOI{\tempurl}


\bibitem[{growtoups}({[n.\,d.]})]%
        {growtoups}
\bibfield{author}{\bibinfo{person}{{growtoups}}.}
  \bibinfo{year}{[n.\,d.]}\natexlab{}.
\newblock \bibinfo{title}{{Datacenter ASN Blocking}}.
\newblock \bibinfo{howpublished}{\url{https://github.com/growtoups/ASN_LIST}}.
\newblock


\bibitem[Hu et~al\mbox{.}(2009)]%
        {rbseeker}
\bibfield{author}{\bibinfo{person}{Xin Hu}, \bibinfo{person}{Matthew Knysz},
  {and} \bibinfo{person}{Kang~G Shin}.} \bibinfo{year}{2009}\natexlab{}.
\newblock \showarticletitle{RB-Seeker: Auto-detection of Redirection Botnets.}.
  In \bibinfo{booktitle}{\emph{NDSS}}.
\newblock


\bibitem[Ikram et~al\mbox{.}(2016)]%
        {oneclass-tracker-detection}
\bibfield{author}{\bibinfo{person}{Muhammad Ikram},
  \bibinfo{person}{Hassan~Jameel Asghar}, \bibinfo{person}{Mohamed~Ali
  K{\^{a}}afar}, \bibinfo{person}{Balachander Krishnamurthy}, {and}
  \bibinfo{person}{Anirban Mahanti}.} \bibinfo{year}{2016}\natexlab{}.
\newblock \showarticletitle{Towards Seamless Tracking-Free Web: Improved
  Detection of Trackers via One-class Learning}.
\newblock \bibinfo{journal}{\emph{CoRR}}  \bibinfo{volume}{abs/1603.06289}
  (\bibinfo{year}{2016}).
\newblock
\showeprint[arXiv]{1603.06289}
\urldef\tempurl%
\url{http://arxiv.org/abs/1603.06289}
\showURL{%
\tempurl}


\bibitem[Iliou et~al\mbox{.}(2019)]%
        {Iliou2019ARES}
\bibfield{author}{\bibinfo{person}{Christos Iliou}, \bibinfo{person}{Theodoros
  Kostoulas}, \bibinfo{person}{Theodora Tsikrika}, \bibinfo{person}{Vasilis
  Katos}, \bibinfo{person}{Stefanos Vrochidis}, {and} \bibinfo{person}{Yiannis
  Kompatsiaris}.} \bibinfo{year}{2019}\natexlab{}.
\newblock \showarticletitle{Towards a Framework for Detecting Advanced Web
  Bots}. In \bibinfo{booktitle}{\emph{Proceedings of the 14th International
  Conference on Availability, Reliability and Security (ARES 2019)}}.
  \bibinfo{publisher}{Association for Computing Machinery},
  \bibinfo{address}{New York, NY, USA}.
\newblock


\bibitem[{imperva.com}({[n.\,d.]})]%
        {imperva_2024}
\bibfield{author}{\bibinfo{person}{{imperva.com}}.}
  \bibinfo{year}{[n.\,d.]}\natexlab{}.
\newblock \bibinfo{title}{{2023 Imperva Bad Bot Report}}.
\newblock
  \bibinfo{howpublished}{\url{https://www.imperva.com/resources/resource-library/reports/2024-bad-bot-report/}}.
\newblock


\bibitem[Inc.({[n.\,d.]})]%
        {AdblockPlus}
\bibfield{author}{\bibinfo{person}{Adblock Inc.}}
  \bibinfo{year}{[n.\,d.]}\natexlab{}.
\newblock \bibinfo{title}{Adblock Plus}.
\newblock
  \bibinfo{howpublished}{\url{https://gitlab.com/adblockinc/ext/adblockplus/adblockplus}}.
\newblock


\bibitem[Iqbal et~al\mbox{.}(2020)]%
        {fp-inspector}
\bibfield{author}{\bibinfo{person}{Umar Iqbal}, \bibinfo{person}{Steven
  Englehardt}, {and} \bibinfo{person}{Zubair Shafiq}.}
  \bibinfo{year}{2020}\natexlab{}.
\newblock \showarticletitle{Fingerprinting the Fingerprinters: Learning to
  Detect Browser Fingerprinting Behaviors}.
\newblock \bibinfo{journal}{\emph{CoRR}}  \bibinfo{volume}{abs/2008.04480}
  (\bibinfo{year}{2020}).
\newblock
\showeprint[arXiv]{2008.04480}
\urldef\tempurl%
\url{https://arxiv.org/abs/2008.04480}
\showURL{%
\tempurl}


\bibitem[Iqbal et~al\mbox{.}(2018)]%
        {adgraph}
\bibfield{author}{\bibinfo{person}{Umar Iqbal}, \bibinfo{person}{Zubair
  Shafiq}, \bibinfo{person}{Peter Snyder}, \bibinfo{person}{Shitong Zhu},
  \bibinfo{person}{Zhiyun Qian}, {and} \bibinfo{person}{Benjamin Livshits}.}
  \bibinfo{year}{2018}\natexlab{}.
\newblock \showarticletitle{AdGraph: {A} Machine Learning Approach to Automatic
  and Effective Adblocking}.
\newblock \bibinfo{journal}{\emph{CoRR}}  \bibinfo{volume}{abs/1805.09155}
  (\bibinfo{year}{2018}).
\newblock
\showeprint[arXiv]{1805.09155}
\urldef\tempurl%
\url{http://arxiv.org/abs/1805.09155}
\showURL{%
\tempurl}


\bibitem[Javed et~al\mbox{.}(2015)]%
        {javed2015measurement}
\bibfield{author}{\bibinfo{person}{Mobin Javed}, \bibinfo{person}{Cormac
  Herley}, \bibinfo{person}{Marcus Peinado}, {and} \bibinfo{person}{Vern
  Paxson}.} \bibinfo{year}{2015}\natexlab{}.
\newblock \showarticletitle{Measurement and analysis of traffic exchange
  services}. In \bibinfo{booktitle}{\emph{Proceedings of the 2015 Internet
  Measurement Conference}}. \bibinfo{pages}{1--12}.
\newblock


\bibitem[Jin et~al\mbox{.}(2013)]%
        {Jin2013IEEE}
\bibfield{author}{\bibinfo{person}{Jing Jin}, \bibinfo{person}{Jeff Offutt},
  \bibinfo{person}{Nan Zheng}, \bibinfo{person}{Feng Mao},
  \bibinfo{person}{Aaron Koehl}, {and} \bibinfo{person}{Haining Wang}.}
  \bibinfo{year}{2013}\natexlab{}.
\newblock \showarticletitle{Evasive Bots Masquerading as Human Beings on the
  Web}. In \bibinfo{booktitle}{\emph{2013 43rd Annual IEEE/IFIP International
  Conference on Dependable Systems and Networks (DSN)}}.
  \bibinfo{publisher}{IEEE}, \bibinfo{address}{New York, NY, USA},
  \bibinfo{pages}{1--12}.
\newblock
\urldef\tempurl%
\url{https://doi.org/10.1109/DSN.2013.6575366}
\showDOI{\tempurl}


\bibitem[Jueckstock et~al\mbox{.}(2021)]%
        {reliable-web-crawling}
\bibfield{author}{\bibinfo{person}{Jordan Jueckstock}, \bibinfo{person}{Shaown
  Sarker}, \bibinfo{person}{Peter Snyder}, \bibinfo{person}{Aidan Beggs},
  \bibinfo{person}{Panagiotis Papadopoulos}, \bibinfo{person}{Matteo Varvello},
  \bibinfo{person}{Benjamin Livshits}, {and} \bibinfo{person}{Alexandros
  Kapravelos}.} \bibinfo{year}{2021}\natexlab{}.
\newblock \showarticletitle{Towards Realistic and ReproducibleWeb Crawl
  Measurements} \emph{(\bibinfo{series}{WWW '21})}.
  \bibinfo{publisher}{Association for Computing Machinery},
  \bibinfo{address}{New York, NY, USA}, \bibinfo{pages}{80–91}.
\newblock
\urldef\tempurl%
\url{https://doi.org/10.1145/3442381.3450050}
\showDOI{\tempurl}


\bibitem[Kalantari et~al\mbox{.}(2024)]%
        {Kalantari2024BrowserPolygraph}
\bibfield{author}{\bibinfo{person}{Faezeh Kalantari},
  \bibinfo{person}{Mehrnoosh Zaeifi}, \bibinfo{person}{Yeganeh Safaei},
  \bibinfo{person}{Marzieh Bitaab}, \bibinfo{person}{Adam Oest},
  \bibinfo{person}{Gianluca Stringhini}, \bibinfo{person}{Yan Shoshitaishvili},
  {and} \bibinfo{person}{Adam Doup\'{e}}.} \bibinfo{year}{2024}\natexlab{}.
\newblock \showarticletitle{Browser Polygraph: Efficient Deployment of
  Coarse-Grained Browser Fingerprints for Web-Scale Detection of Fraud
  Browsers}. In \bibinfo{booktitle}{\emph{Proceedings of the 2024 ACM on
  Internet Measurement Conference}}.
\newblock
\urldef\tempurl%
\url{https://doi.org/10.1145/3646547.3688455}
\showURL{%
\tempurl}


\bibitem[Laor et~al\mbox{.}(2022)]%
        {drawn_apart-gpu}
\bibfield{author}{\bibinfo{person}{Tomer Laor}, \bibinfo{person}{Naif Mehanna},
  \bibinfo{person}{Antonin Durey}, \bibinfo{person}{Vitaly Dyadyuk},
  \bibinfo{person}{Pierre Laperdrix}, \bibinfo{person}{Cl{\'{e} }mentine
  Maurice}, \bibinfo{person}{Yossi Oren}, \bibinfo{person}{Romain Rouvoy},
  \bibinfo{person}{Walter Rudametkin}, {and} \bibinfo{person}{Yuval Yarom}.}
  \bibinfo{year}{2022}\natexlab{}.
\newblock \showarticletitle{{DRAWN} {APART} : A Device Identification Technique
  based on Remote {GPU} Fingerprinting}. In
  \bibinfo{booktitle}{\emph{Proceedings 2022 Network and Distributed System
  Security Symposium}}. \bibinfo{publisher}{Internet Society}.
\newblock
\urldef\tempurl%
\url{https://doi.org/10.14722/ndss.2022.24093}
\showDOI{\tempurl}


\bibitem[Li et~al\mbox{.}(2021)]%
        {Li2021S&P}
\bibfield{author}{\bibinfo{person}{Xigao Li}, \bibinfo{person}{Babak~Amin
  Azad}, \bibinfo{person}{Amir Rahmati}, {and} \bibinfo{person}{Nick
  Nikiforakis}.} \bibinfo{year}{2021}\natexlab{}.
\newblock \showarticletitle{Good Bot, Bad Bot: Characterizing Automated
  Browsing Activity}. In \bibinfo{booktitle}{\emph{2021 IEEE Symposium on
  Security and Privacy (SP)}}. \bibinfo{pages}{1589--1605}.
\newblock
\urldef\tempurl%
\url{https://doi.org/10.1109/SP40001.2021.00079}
\showDOI{\tempurl}


\bibitem[Liu et~al\mbox{.}(2022)]%
        {liu2022gummy}
\bibfield{author}{\bibinfo{person}{Zengrui Liu}, \bibinfo{person}{Prakash
  Shrestha}, {and} \bibinfo{person}{Nitesh Saxena}.}
  \bibinfo{year}{2022}\natexlab{}.
\newblock \showarticletitle{Gummy browsers: targeted browser spoofing against
  state-of-the-art fingerprinting techniques}. In
  \bibinfo{booktitle}{\emph{International Conference on Applied Cryptography
  and Network Security}}. Springer, \bibinfo{pages}{147--169}.
\newblock


\bibitem[{MaxMind}(2024a)]%
        {maxmind-geoip}
\bibfield{author}{\bibinfo{person}{{MaxMind}}.}
  \bibinfo{year}{2024}\natexlab{a}.
\newblock \bibinfo{title}{{MaxMind GeoIP Databases}}.
\newblock
  \bibinfo{howpublished}{\url{https://www.maxmind.com/en/geoip-databases}}.
\newblock


\bibitem[{MaxMind}(2024b)]%
        {maxmind-minfraud}
\bibfield{author}{\bibinfo{person}{{MaxMind}}.}
  \bibinfo{year}{2024}\natexlab{b}.
\newblock \bibinfo{title}{{MaxMind minFraud Services}}.
\newblock
  \bibinfo{howpublished}{\url{https://www.maxmind.com/en/solutions/fraud-prevention/overview}}.
\newblock


\bibitem[{Mozilla}({[n.\,d.]})]%
        {get-timezone-offset}
\bibfield{author}{\bibinfo{person}{{Mozilla}}.}
  \bibinfo{year}{[n.\,d.]}\natexlab{}.
\newblock \bibinfo{title}{{Date.prototype.getTimezoneOffset()}}.
\newblock
  \bibinfo{howpublished}{\url{https://develper.mozilla.org/enUS/docs/Web/JavaScript/Reference/Global_Objects/Date/getTimezoneOffset}}.
\newblock


\bibitem[Munir et~al\mbox{.}(2023)]%
        {cookiegraph}
\bibfield{author}{\bibinfo{person}{Shaoor Munir}, \bibinfo{person}{Sandra
  Siby}, \bibinfo{person}{Umar Iqbal}, \bibinfo{person}{Steven Englehardt},
  \bibinfo{person}{Zubair Shafiq}, {and} \bibinfo{person}{Carmela Troncoso}.}
  \bibinfo{year}{2023}\natexlab{}.
\newblock \bibinfo{title}{COOKIEGRAPH: Understanding and Detecting First-Party
  Tracking Cookies}.
\newblock
\newblock
\showeprint[arxiv]{2208.12370}~[cs.CR]


\bibitem[Nakatsuka et~al\mbox{.}(2021)]%
        {cacti}
\bibfield{author}{\bibinfo{person}{Yoshimichi Nakatsuka},
  \bibinfo{person}{Ercan Ozturk}, \bibinfo{person}{Andrew Paverd}, {and}
  \bibinfo{person}{Gene Tsudik}.} \bibinfo{year}{2021}\natexlab{}.
\newblock \showarticletitle{{CACTI}: Captcha Avoidance via Client-side {TEE}
  Integration}. In \bibinfo{booktitle}{\emph{30th USENIX Security Symposium
  (USENIX Security 21)}}. \bibinfo{publisher}{USENIX Association},
  \bibinfo{pages}{2561--2578}.
\newblock
\showISBNx{978-1-939133-24-3}
\urldef\tempurl%
\url{https://www.usenix.org/conference/usenixsecurity21/presentation/nakatsuka}
\showURL{%
\tempurl}


\bibitem[Nguyen~Ba et~al\mbox{.}(2021)]%
        {csci_casestudy}
\bibfield{author}{\bibinfo{person}{Minh~Hieu Nguyen~Ba}, \bibinfo{person}{Jacob
  Bennett}, \bibinfo{person}{Michael Gallagher}, {and} \bibinfo{person}{Suman
  Bhunia}.} \bibinfo{year}{2021}\natexlab{}.
\newblock \showarticletitle{A Case Study of Credential Stuffing Attack: Canva
  Data Breach}. In \bibinfo{booktitle}{\emph{2021 International Conference on
  Computational Science and Computational Intelligence (CSCI)}}.
  \bibinfo{pages}{735--740}.
\newblock
\urldef\tempurl%
\url{https://doi.org/10.1109/CSCI54926.2021.00187}
\showDOI{\tempurl}


\bibitem[{OpenWPM}({[n.\,d.]})]%
        {openwpm}
\bibfield{author}{\bibinfo{person}{{OpenWPM}}.}
  \bibinfo{year}{[n.\,d.]}\natexlab{}.
\newblock \bibinfo{title}{{A web privacy measurement framework}}.
\newblock \bibinfo{howpublished}{\url{https://github.com/openwpm/OpenWPM}}.
\newblock


\bibitem[Sanchez-Rola et~al\mbox{.}(2018)]%
        {clock-skew}
\bibfield{author}{\bibinfo{person}{Iskander Sanchez-Rola},
  \bibinfo{person}{Igor Santos}, {and} \bibinfo{person}{Davide Balzarotti}.}
  \bibinfo{year}{2018}\natexlab{}.
\newblock \showarticletitle{{Clock Around the Clock: Time-Based Device
  Fingerprinting}}. In \bibinfo{booktitle}{\emph{Proceedings of the 2018 ACM
  SIGSAC Conference on Computer and Communications Security}} (Toronto, Canada)
  \emph{(\bibinfo{series}{CCS '18})}. \bibinfo{publisher}{Association for
  Computing Machinery}, \bibinfo{address}{New York, NY, USA},
  \bibinfo{pages}{1502–1514}.
\newblock
\showISBNx{9781450356930}
\urldef\tempurl%
\url{https://doi.org/10.1145/3243734.3243796}
\showDOI{\tempurl}


\bibitem[Schaller et~al\mbox{.}(2017)]%
        {rowhammer-puf}
\bibfield{author}{\bibinfo{person}{Andre Schaller}, \bibinfo{person}{Wenjie
  Xiong}, \bibinfo{person}{Nikolaos~Athanasios Anagnostopoulos},
  \bibinfo{person}{Muhammad~Umair Saleem}, \bibinfo{person}{Sebastian
  Gabmeyer}, \bibinfo{person}{Stefan Katzenbeisser}, {and}
  \bibinfo{person}{Jakub Szefer}.} \bibinfo{year}{2017}\natexlab{}.
\newblock \showarticletitle{Intrinsic Rowhammer {PUFs}: Leveraging the
  Rowhammer effect for improved security}. In \bibinfo{booktitle}{\emph{2017
  {IEEE} International Symposium on Hardware Oriented Security and Trust
  ({HOST})}}. \bibinfo{publisher}{{IEEE}}.
\newblock
\urldef\tempurl%
\url{https://doi.org/10.1109/hst.2017.7951729}
\showDOI{\tempurl}


\bibitem[Searles et~al\mbox{.}(2023)]%
        {captcha_annoyance}
\bibfield{author}{\bibinfo{person}{Andrew Searles}, \bibinfo{person}{Yoshimichi
  Nakatsuka}, \bibinfo{person}{Ercan Ozturk}, \bibinfo{person}{Andrew Paverd},
  \bibinfo{person}{Gene Tsudik}, {and} \bibinfo{person}{Ai Enkoji}.}
  \bibinfo{year}{2023}\natexlab{}.
\newblock \showarticletitle{An Empirical Study \& Evaluation of Modern
  {CAPTCHAs}}. In \bibinfo{booktitle}{\emph{32nd USENIX Security Symposium
  (USENIX Security 23)}}. \bibinfo{publisher}{USENIX Association},
  \bibinfo{address}{Anaheim, CA}, \bibinfo{pages}{3081--3097}.
\newblock
\showISBNx{978-1-939133-37-3}
\urldef\tempurl%
\url{https://www.usenix.org/conference/usenixsecurity23/presentation/searles}
\showURL{%
\tempurl}


\bibitem[{seoclerks}({[n.\,d.]})]%
        {seoclerks}
\bibfield{author}{\bibinfo{person}{{seoclerks}}.}
  \bibinfo{year}{[n.\,d.]}\natexlab{}.
\newblock \bibinfo{title}{{SEO Marketplace for backlinks, web design, website
  traffic, and online marketing}}.
\newblock \bibinfo{howpublished}{\url{https://www.seoclerks.com/}}.
\newblock


\bibitem[{SHapley Additive exPlanations}({[n.\,d.]})]%
        {shap}
\bibfield{author}{\bibinfo{person}{{SHapley Additive exPlanations}}.}
  \bibinfo{year}{[n.\,d.]}\natexlab{}.
\newblock \bibinfo{title}{{Welcome to the SHAP documentation}}.
\newblock \bibinfo{howpublished}{\url{https://shap.readthedocs.io/en/latest/}}.
\newblock


\bibitem[Siby et~al\mbox{.}(2022)]%
        {webgraph}
\bibfield{author}{\bibinfo{person}{Sandra Siby}, \bibinfo{person}{Umar Iqbal},
  \bibinfo{person}{Steven Englehardt}, \bibinfo{person}{Zubair Shafiq}, {and}
  \bibinfo{person}{Carmela Troncoso}.} \bibinfo{year}{2022}\natexlab{}.
\newblock \showarticletitle{{WebGraph}: Capturing Advertising and Tracking
  Information Flows for Robust Blocking}. In \bibinfo{booktitle}{\emph{31st
  USENIX Security Symposium (USENIX Security 22)}}. \bibinfo{publisher}{USENIX
  Association}, \bibinfo{address}{Boston, MA}, \bibinfo{pages}{2875--2892}.
\newblock
\showISBNx{978-1-939133-31-1}
\urldef\tempurl%
\url{https://www.usenix.org/conference/usenixsecurity22/presentation/siby}
\showURL{%
\tempurl}


\bibitem[{Spark Traffic}({[n.\,d.]})]%
        {spark}
\bibfield{author}{\bibinfo{person}{{Spark Traffic}}.}
  \bibinfo{year}{[n.\,d.]}\natexlab{}.
\newblock \bibinfo{title}{{Comprehensive Marketing Suite for better SEO
  ranking}}.
\newblock \bibinfo{howpublished}{\url{https://www.sparktraffic.com/}}.
\newblock


\bibitem[Springborn and Barford(2013)]%
        {impression-fraud}
\bibfield{author}{\bibinfo{person}{Kevin Springborn} {and}
  \bibinfo{person}{Paul Barford}.} \bibinfo{year}{2013}\natexlab{}.
\newblock \showarticletitle{Impression Fraud in On-line Advertising via
  {Pay-Per-View} Networks}. In \bibinfo{booktitle}{\emph{22nd USENIX Security
  Symposium (USENIX Security 13)}}. \bibinfo{publisher}{USENIX Association},
  \bibinfo{address}{Washington, D.C.}, \bibinfo{pages}{211--226}.
\newblock
\showISBNx{978-1-931971-03-4}
\urldef\tempurl%
\url{https://www.usenix.org/conference/usenixsecurity13/technical-sessions/paper/springborn}
\showURL{%
\tempurl}


\bibitem[StatCounter(2024)]%
        {browser_market}
\bibfield{author}{\bibinfo{person}{StatCounter}.}
  \bibinfo{year}{2024}\natexlab{}.
\newblock \bibinfo{title}{Browser Market Share Worldwide}.
\newblock
  \bibinfo{howpublished}{\url{https://gs.statcounter.com/browser-market-share}}.
\newblock


\bibitem[TechReport({[n.\,d.]})]%
        {brave-market-share}
\bibfield{author}{\bibinfo{person}{TechReport}.}
  \bibinfo{year}{[n.\,d.]}\natexlab{}.
\newblock \bibinfo{title}{Most Important Brave Market Share Statistics in
  2024}.
\newblock
  \bibinfo{howpublished}{\url{https://techreport.com/statistics/software-web/brave-market-share-statistics/}}.
\newblock


\bibitem[Tor({[n.\,d.]})]%
        {tor-block}
\bibfield{author}{\bibinfo{person}{Tor}.} \bibinfo{year}{[n.\,d.]}\natexlab{}.
\newblock \bibinfo{title}{A website I am trying to reach is blocking access
  over Tor.}
\newblock
  \bibinfo{howpublished}{\url{https://support.torproject.org/tbb/website-blocking-tor/}}.
\newblock


\bibitem[{Tor}({[n.\,d.]})]%
        {tor-browser}
\bibfield{author}{\bibinfo{person}{{Tor}}.}
  \bibinfo{year}{[n.\,d.]}\natexlab{}.
\newblock \bibinfo{title}{{You have a right to BROWSE without being watched}}.
\newblock
  \bibinfo{howpublished}{\url{https://www.torproject.org/download/languages/}}.
\newblock


\bibitem[{U.S. Department of Health and Human Services}(2018)]%
        {HHSDecisionCharts2018}
\bibfield{author}{\bibinfo{person}{{U.S. Department of Health and Human
  Services}}.} \bibinfo{year}{2018}\natexlab{}.
\newblock \bibinfo{title}{Decision Charts: 2018 Requirements (Common Rule)}.
\newblock
\newblock
\urldef\tempurl%
\url{https://www.hhs.gov/ohrp/regulations-and-policy/decision-charts-2018/index.html#c1}
\showURL{%
\tempurl}


\bibitem[Vastel et~al\mbox{.}(2018)]%
        {Vastel2018SP}
\bibfield{author}{\bibinfo{person}{Antoine Vastel}, \bibinfo{person}{Pierre
  Laperdrix}, \bibinfo{person}{Walter Rudametkin}, {and}
  \bibinfo{person}{Romain Rouvoy}.} \bibinfo{year}{2018}\natexlab{}.
\newblock \showarticletitle{FP-STALKER: Tracking Browser Fingerprint
  Evolutions}. In \bibinfo{booktitle}{\emph{2018 IEEE Symposium on Security and
  Privacy (SP)}}. \bibinfo{pages}{728--741}.
\newblock
\urldef\tempurl%
\url{https://doi.org/10.1109/SP.2018.00008}
\showDOI{\tempurl}


\bibitem[Vastel et~al\mbox{.}(2020)]%
        {Vastel2020MADWeb}
\bibfield{author}{\bibinfo{person}{Antoine Vastel}, \bibinfo{person}{Walter
  Rudametkin}, \bibinfo{person}{Romain Rouvoy}, {and} \bibinfo{person}{Xavier
  Blanc}.} \bibinfo{year}{2020}\natexlab{}.
\newblock \showarticletitle{{FP-Crawlers: Studying the Resilience of Browser
  Fingerprinting to Block Crawlers}}. In \bibinfo{booktitle}{\emph{{MADWeb'20 -
  NDSS Workshop on Measurements, Attacks, and Defenses for the Web}}},
  \bibfield{editor}{\bibinfo{person}{Oleksii Starov},
  \bibinfo{person}{Alexandros Kapravelos}, {and} \bibinfo{person}{Nick
  Nikiforakis}} (Eds.). \bibinfo{address}{San Diego, United States}.
\newblock
\urldef\tempurl%
\url{https://doi.org/10.14722/ndss.2020.23xxx}
\showDOI{\tempurl}


\bibitem[Venugopalan et~al\mbox{.}(2023)]%
        {centauri}
\bibfield{author}{\bibinfo{person}{Hari Venugopalan}, \bibinfo{person}{Kaustav
  Goswami}, \bibinfo{person}{Zainul~Abi Din}, \bibinfo{person}{Jason
  Lowe-Power}, \bibinfo{person}{Samuel~T. King}, {and} \bibinfo{person}{Zubair
  Shafiq}.} \bibinfo{year}{2023}\natexlab{}.
\newblock \bibinfo{title}{Centauri: Practical Rowhammer Fingerprinting}.
\newblock
\newblock
\showeprint[arxiv]{2307.00143}~[cs.CR]


\bibitem[Webstore(2024a)]%
        {adblock_webstore}
\bibfield{author}{\bibinfo{person}{Chrome Webstore}.}
  \bibinfo{year}{2024}\natexlab{a}.
\newblock \bibinfo{title}{Adblock Plus}.
\newblock
  \bibinfo{howpublished}{\url{https://chromewebstore.google.com/detail/adblock-plus-free-ad-bloc/cfhdojbkjhnklbpkdaibdccddilifddb}}.
\newblock


\bibitem[Webstore(2024b)]%
        {ublock_webstore}
\bibfield{author}{\bibinfo{person}{Chrome Webstore}.}
  \bibinfo{year}{2024}\natexlab{b}.
\newblock \bibinfo{title}{uBlock Origin}.
\newblock
  \bibinfo{howpublished}{\url{https://chromewebstore.google.com/detail/ublock-origin/cjpalhdlnbpafiamejdnhcphjbkeiagm}}.
\newblock


\bibitem[Wu et~al\mbox{.}(2023)]%
        {him-many-faces}
\bibfield{author}{\bibinfo{person}{Shujiang Wu}, \bibinfo{person}{Pengfei Sun},
  \bibinfo{person}{Yao Zhao}, {and} \bibinfo{person}{Yinzhi Cao}.}
  \bibinfo{year}{2023}\natexlab{}.
\newblock \showarticletitle{Him of Many Faces: Characterizing Billion-scale
  Adversarial and Benign Browser Fingerprints on Commercial Websites}. In
  \bibinfo{booktitle}{\emph{30th Annual Network and Distributed System Security
  Symposium, {NDSS} 2023, San Diego, California, USA, February 27 - March 3,
  2023}}. \bibinfo{publisher}{The Internet Society}.
\newblock


\bibitem[{XGBoost}({[n.\,d.]})]%
        {xgboost}
\bibfield{author}{\bibinfo{person}{{XGBoost}}.}
  \bibinfo{year}{[n.\,d.]}\natexlab{}.
\newblock \bibinfo{title}{{XGBoost Documentation}}.
\newblock
  \bibinfo{howpublished}{\url{https://xgboost.readthedocs.io/en/stable/}}.
\newblock


\bibitem[Yasuhara et~al\mbox{.}(2024)]%
        {datadome_citation_paper}
\bibfield{author}{\bibinfo{person}{Kazuki Yasuhara}, \bibinfo{person}{Naoki
  Kodama}, {and} \bibinfo{person}{Takamichi Saito}.}
  \bibinfo{year}{2024}\natexlab{}.
\newblock \showarticletitle{Challenges in Web Bot Detection and Detection
  Evasion Technologies}. In \bibinfo{booktitle}{\emph{Advances in Network-Based
  Information Systems}}, \bibfield{editor}{\bibinfo{person}{Leonard Barolli}}
  (Ed.). \bibinfo{publisher}{Springer Nature Switzerland},
  \bibinfo{address}{Cham}, \bibinfo{pages}{162--173}.
\newblock
\showISBNx{978-3-031-72325-4}


\end{thebibliography}

\appendix
\section{Ethics}
\label{app:ethics}
This study complies with ethical guidelines for research involving data collection and usage.
To conduct the research, we paid a small amount to bot operators to generate requests directed solely to our honey site.
To ensure data quality and realism, we prioritized bot services with high ratings and traffic advertised as realistic and organic.
These requests were analyzed exclusively for research purposes, with the goal of improving bot detection.

The research process was reviewed and approved by our university, ensuring alignment with ethical principles outlined in both the Belmont Report and the Menlo Report.
To determine whether Institutional Review Board (IRB) approval was necessary, we consulted official guidelines from the Human Subject Regulations Decision Charts \cite{HHSDecisionCharts2018}, specifically the section addressing activities covered by 45 CFR Part 46.
Based on this evaluation, we determined that our research does not involve human subjects as defined by 45 CFR Part 46 and, as a result, qualifies for exemption from IRB oversight.

Furthermore, our study did not collect or store any Personally Identifiable Information (PII), nor did it involve the identification or tracking of individual users across different websites/contexts. 
Traffic data was analyzed in aggregate, and identifiable information, such as IP addresses, was hashed before storage.

All purchased traffic was directed exclusively towards our honey site, ensuring that no other sites or users were impacted.
The primary purpose of this research was to advance the science of bot detection, and we refrained from monetizing the honey site or deriving any profit from the generated traffic.

\section{Comparison of APIs used by \botd and \datadome}
\label{app:api-comparison}
Table \ref{tab:api_comparison_botd_datadome} shows the different APIs accessed by \botd and \datadome scripts on our honey site.

\begin{table}[!htpb]
   \centering
   \caption{Comparison of browser APIs read by DataDome and BotD}
   
    \resizebox*{\linewidth}{!}{\begin{tabular}{l c c}
      \toprule
      \textbf{Browser API}                 & \textbf{DataDome} & \textbf{BotD} \\
      \midrule
      \multicolumn{3}{c}{\textbf{Display}}                                     \\
       \midrule
          window.screen.colorDepth                & \tikzcmark & \tikzxmark \\
          HTMLCanvasElement.getContext            & \tikzcmark & \tikzcmark \\
        \midrule
         \multicolumn{3}{c}{\textbf{Navigator}} \\
         \midrule
          window.navigator.webdriver              & \tikzcmark & \tikzcmark \\
          window.navigator.vendor                 & \tikzcmark & \tikzcmark \\
          window.navigator.userAgent              & \tikzcmark & \tikzcmark \\
          window.navigator.serviceWorker          & \tikzcmark & \tikzxmark \\
          window.navigator.productSub             & \tikzxmark & \tikzcmark \\
          window.navigator.plugins                & \tikzcmark & \tikzcmark \\
          window.navigator.platform               & \tikzcmark & \tikzxmark \\
          window.navigator.permissions            & \tikzcmark & \tikzcmark \\
          window.navigator.oscpu                  & \tikzcmark & \tikzxmark \\
          window.navigator.mimeTypes              & \tikzcmark & \tikzcmark \\
          window.navigator.mediaDevices           & \tikzcmark & \tikzxmark \\
          window.navigator.maxTouchPoints         & \tikzcmark & \tikzxmark \\
          window.navigator.languages              & \tikzcmark & \tikzcmark \\
          window.navigator.language               & \tikzcmark & \tikzcmark \\
          window.navigator.hardwareConcurrency    & \tikzcmark & \tikzxmark \\
          window.navigator.buildID                & \tikzcmark & \tikzxmark \\
          window.navigator.appVersion             & \tikzxmark & \tikzcmark \\
          window.navigator.\_\_proto\_\_              & \tikzcmark & \tikzxmark \\
        \midrule
         \multicolumn{3}{c}{\textbf{Storage}} \\
         \midrule
          window.sessionStorage                   & \tikzcmark & \tikzxmark \\
          window.localStorage                     & \tikzcmark & \tikzxmark \\
          window.document.cookie                  & \tikzcmark & \tikzxmark \\
        \midrule
         \multicolumn{3}{c}{\textbf{Mouse Movements}} \\
         \midrule
          MouseEvent.type                         & \tikzcmark & \tikzxmark \\
          MouseEvent.timeStamp                    & \tikzcmark & \tikzxmark \\
          MouseEvent.clientY                      & \tikzcmark & \tikzxmark \\
          MouseEvent.clientX                      & \tikzcmark & \tikzxmark \\
          addEventListner: mouseup                & \tikzcmark & \tikzxmark \\
          addEventListner: mousemove              & \tikzcmark & \tikzxmark \\
          addEventListner: mousedown              & \tikzcmark & \tikzxmark \\
          \midrule
         \multicolumn{3}{c}{\textbf{Miscellaneous}} \\
         \midrule
          addEventListner: asyncChallengeFinished & \tikzcmark & \tikzxmark \\
          addEventListner: pagehide               & \tikzcmark & \tikzxmark \\
          Performance.now                         & \tikzcmark & \tikzxmark \\
          \bottomrule
   \end{tabular}}
   \label{tab:api_comparison_botd_datadome}
\end{table}

\section{Combination of fingerprint attributes to evade \datadome}
\label{app:datadome_tree}
We visualized the XGBoost decision tree for \datadome described in Section \ref{sec:measurement_attributes}. The tree with a depth of 5 indicated that
all 44,168 requests having a Screen Frame value less than 20 that do not support
the Chrome PDF Viewer plugin, having memory over 256 MB with less than
14 CPU cores having the width of Monospace font used in FingerprintJS
larger than 131.5 were able to evade detection.

\begin{algorithm}
\caption{Algorithm to Detect Spatial Inconsistencies}
\label{alg:spatial_inconsistencies}
\begin{algorithmic}[1]
    \STATE \textbf{Input:} Attribute categories $F$, Dataset containing requests $D$, Labels for requests $L$ (\textit{true} if the request is from a bot, \textit{false} for human)
    \FORALL{$f \in F$}
        \FORALL{attribute pairs $\{f_a, f_b\} \subseteq f$}
            \STATE Filter $D$ where $L$ is \textit{false}, creating $D'$
            \STATE Create tuples $(v_{f_a}, nv_{f_b})$, where $v_{f_a}$ is the value of $f_a$ and $nv_{f_b}$ is the number of unique values of $f_b$ found in the same row as $v_{f_a}$ in $D'$
            \STATE Sort the tuples in increasing order of $nv_{f_b}$
            \FORALL{$(v_{f_a}, nv_{f_b})$ in the sorted order}
                \IF{the combination is inconsistent}
                    \STATE Label all rows in $D$ containing $(v_{f_a}, v_{f_b})$ as \textit{true}
                \ENDIF
            \ENDFOR
        \ENDFOR
    \ENDFOR
\end{algorithmic}
\end{algorithm}

\section{Algorithm to identify spatial inconsistencies}
\label{app:alg}
Algorithm \ref{alg:spatial_inconsistencies} describes our algorithm to
identify spatial inconsistencies.

\begin{table*}[!htpb]
\centering

\caption{Inconsistencies Identified}
\begin{tabular}{l|l|l}
\toprule
\textbf{Attribute Group} & \textbf{Attributes} & \textbf{Examples} \\
\midrule

\multirow{15}{*}{Screen} & \multirow{5}{*}{(UA Device, Screen Resolution)} & \texttt{(iPhone, 1920x1080)} \\
& & \texttt{(iPhone, 847x476)} \\
& & \texttt{(iPad, 900x1600)} \\
& & \texttt{(Samsung SM-S906N, 1920x1080)} \\
& & \texttt{(M2006C3MG, 800x360)} \\
& & \texttt{(Mac, 656x1364)} \\
\cline{2-3}
& \multirow{5}{*}{(UA Device, Touch Support)} & \texttt{(iPhone, None)} \\
& & \texttt{(Mac, touchEvent/touchStart)} \\
& & \texttt{(Samsung SM-A127F, None)} \\
& & \texttt{(M2004J19C, None)} \\
& & \texttt{(Infinix X652B, None)} \\
\cline{2-3}
& \multirow{5}{*}{(UA Device, Max Touch Points)} & \texttt{(iPhone, 1)} \\
& & \texttt{(iPhone, 0)} \\
& & \texttt{(iPad, 1)} \\
& & \texttt{(iPad, 7)} \\
& & \texttt{(Mac, 10)} \\
& & \texttt{(Samsung SM-A515F, 0)} \\
& & \texttt{(Pixel 7 Pro, 0)} \\[0.5ex]
\cline{2-3}
& \multirow{2}{*}{(UA Device, Color Depth)} & \texttt{(iPhone, 16)} \\
& & \texttt{(iPad, 16)} \\
\cline{2-3}
& \multirow{2}{*}{(UA Device,  Color Gamut)} & \texttt{(Samsung Galaxy Tab S7, (p3, rec2020))} \\
& & \texttt{(SAM Galaxy S10 Smartphone, (p3, rec2020))} \\
\midrule

\multirow{10}{*}{Device} & \multirow{5}{*}{(UA Device, Device Memory)} & \texttt{(XiaoMi Mi Pad4 LTE, 8)} \\
& & \texttt{(Samsung SM-T387W, 4)} \\
& & \texttt{(MiPad 3, 8)} \\
& & \texttt{(Samsung SM-A515F, 1)} \\
& & \texttt{(XiaoMi Redmi Go, 8)} \\
\cline{2-3}
& \multirow{5}{*}{(UA Device, Hardware Concurrency)} & \texttt{(iPhone, 3)} \\
& & \texttt{(iPhone, 32)} \\
& & \texttt{(Mac, 48)} \\
& & \texttt{(iPad, 32)} \\
& & \texttt{(XiaoMi Mi Pad5 Wi-Fi, 1)} \\
& & \texttt{(Pixel 2, 32)} \\
\midrule

\multirow{9}{*}{Browser} & \multirow{3}{*}{(UA Browser, UA OS)} & \texttt{(Safari, Linux)} \\
& & \texttt{(Samsung Internet, Linux)} \\
& & \texttt{(MiuiBrowser, Linux)} \\
& & \texttt{(Safari, Windows)} \\
\cline{2-3}
& \multirow{2}{*}{(UA Browser, Vendor)} & \texttt{(Mobile Safari, Google Inc.)} \\
& & \texttt{(Chrome Mobile, Apple Computer, Inc.)} \\
\cline{2-3}
& \multirow{4}{*}{(UA Browser, Platform)} & \texttt{(Mobile Safari, Linux x86\_64)} \\
& & \texttt{(Chrome Mobile, Win32)} \\
& & \texttt{(Chrome Mobile, Linux x86\_64)} \\
& & \texttt{(Chrome Mobile iOS, Win32)} \\
\midrule
\multirow{5}{*}{Location} & \multirow{5}{*}{(IP Location, Time Zone)} & \texttt{(France/Hauts-de-France, America/Los Angeles)} \\[0.5ex]
 &  & \texttt{(Germany/Sachsen, America/Los Angeles)} \\[0.5ex]
 &  & \texttt{(Singapore/Singapore, America/Los Angeles)} \\[0.5ex]
 &  & \texttt{(United States of America/California, Asia/Shanghai)} \\[0.5ex]
 &  & \texttt{(United States of America/Virginia, Pacific/Auckland)} \\[0.5ex]
 \midrule
  \multirow{5}{*}{Browser} & \multirow{3}{*}{(Platform, Vendor)} & \texttt{(Linux armv5tejl,  Apple Computer, Inc)} \\[0.5ex]
 &  & \texttt{(Linux aarch64,  Apple Computer, Inc.)} \\[0.5ex]
 &  & \texttt{(Linux armv6l,  Apple Computer, Inc.)} \\[0.5ex]
  &  & \texttt{(Win32,  Apple Computer, Inc.)} \\[0.5ex]
    &  & \texttt{(Linux armv8l,  Apple Computer, Inc.)} \\[0.5ex]
 \cline{2-3}
& \multirow{4}{*}{(Platform, UA OS)} & \texttt{(Mobile Safari, Linux x86\_64)} \\
& & \texttt{(Linux armv8l,  Mac OS X)} \\
& & \texttt{(iPad,  Android)} \\
& & \texttt{(Chrome Mobile iOS, Win32)} \\
& & \texttt{(Linux i686, Mac OS X)} \\
\bottomrule
\end{tabular}
\label{table:inconsistencies_identified}
\end{table*}





\section{Inconsistencies Identified}
\label{app:inconsistencies_identified}
Table \ref{table:inconsistencies_identified} lists some examples of the inconsistencies that we identified for each attribute group in Table \ref{tab:features_by_category}.

\begin{table*}[h] 
\centering
\caption{Attribute Categories}
\label{tab:features_by_category}
 \resizebox*{\linewidth}{!}{\begin{tabularx}{\textwidth}{@{}lXl@{}}
\toprule
\textbf{Category} & \textbf{Attributes} \\
\midrule
\textbf{Screen} & UA Device, Color Depth, Screen Resolution, Touch Support, Max Touch Points, HDR, Contrast, Reduced Motion \\
\textbf{Device} & UA Device, Device Memory, Hardware Concurrency, UA OS\\
\textbf{Browser} & UA Browser, Plugins, Platform, UA OS, UA Vendor, Vendor, Vendor Flavors \\
\textbf{Location} & IP Location, Timezone, Languages \\
\bottomrule
\end{tabularx}}
\end{table*}

\section{Attribute Categories for Inconsistency analysis}
\label{app:inconsistency_groups}
Table \ref{tab:features_by_category} list different categories of attributes used for inconsistency analysis.

\section{\datadome and \botd on Brave and Tor traffic}
\label{app:tor_brave_dd_botd}
Roughly after the first 10 requests on each device, \datadome starts detecting all requests from Brave as bots resulting in a false positive rate of 41\% on the 300 requests described in Section \ref{sec:pet_eval}. \botd on the other hand does not detect any requests as bots.

Similar to \system, \datadome detects all requests from Tor browser as bots while \botd does
not detect any requests as bots. This further sheds light on the difficulty in distinguishing between Tor and bot traffic.
\end{document}